\documentclass[graybox]{svmult}
\usepackage{mathptmx}       
\usepackage{helvet}         
\usepackage{courier}        
\usepackage{type1cm}        
\usepackage{makeidx}         
\usepackage{graphicx}        
\usepackage{multicol}        
\usepackage[bottom]{footmisc}
\makeindex             
\begin{document}

\title{Simulation of dimensionality effects in thermal transport}
\author{Davide Donadio}
\institute{Davide Donadio \at Department of Chemistry, University of California Davis, One Shields Avenue, Davis, CA, 95616; Donostia International Physics Center, 20018, Donostia, Spain; IKERBASQUE, Basque Foundation for Science, Bilbao, Spain; Max Planck Institute for Polymer Research, Ackermannweg 10, 55128 Mainz, Germany \email{ddonadio@ucdavis.edu}
}
%
%
\maketitle


\abstract{The discovery of nanostructures and the development of growth and fabrication techniques of one- and two-dimensional materials provide the possibility to probe experimentally heat transport in low-dimensional systems. Nevertheless measuring the thermal conductivity of these systems is extremely challenging and subject to large uncertainties, thus hindering the chance for a direct comparison between experiments and statistical physics models. Atomistic simulations of realistic nanostructures provide the ideal bridge between abstract models and experiments. After briefly introducing the state of the art of heat transport measurement in nanostructures, and numerical techniques to simulate realistic systems at atomistic level, we review the contribution of lattice dynamics and molecular dynamics simulation to understanding nanoscale thermal transport in systems with reduced dimensionality. We focus on the effect of dimensionality in determining the phononic properties of carbon and semiconducting nanostructures, specifically considering the cases of carbon nanotubes, graphene and of silicon nanowires and ultra-thin membranes, underlying analogies and differences with abstract lattice models.
}

\section{Introduction}
\label{sec:intro7}

The rise of nanoscience, starting with the discovery of C$_{60}$ and other carbon fullerenes \cite{Kroto:1985uw} in the 1980s,  and of carbon nanotubes \cite{Iijima:1991wj} and semiconducting nanostructures \cite{Alivisatos:1996uj} in the 1990s, provided suitable platforms to probe experimentally the physical properties of systems with reduced dimensionality. As expected from theoretical predictions, low-dimensional nanostructures display very different electronic properties from their three-dimensional bulk counterparts. One of the most striking examples was the direct observation of Dirac's cones in graphene by x-ray diffraction \cite{Sprinkle:2009jg}.
In general, one expects very different density of electronic states depending on the dimensionality of the systems, which can be exploited for specific application, for example to enhance the efficiency of thermoelectric energy conversion \cite{HICKS:1993p3198,Dresselhaus:2007hx}. On the other hand such direct consequences of quantum confinement become appreciable only when the confined dimension of the nanostructure is reduced below a certain threshold, typically of the order of few nanometers, which may be difficult to attain \cite{Zhao:2004in}.

In non-metallic systems the main heat carriers are phonons, i.e. quantized lattice vibrations. Analogously to the electronic structure, also the phononic properties of materials and, as a consequence, their thermal properties are deeply affected by dimensionality reduction \cite{Ziman1960}. The symmetry and dimensionality of nanostructures determine their phonon density of states, dispersion relations and the selection rules for scattering processes, thus impacting their heat capacity and their thermal conductivity. However, the way in which dimensionality affects thermal conductivity has not been fully clarified. On the one hand, following the predictions for non-linear models , e.g. the Fermi-Pasta-Ulam model, one would expect that the thermal conductivity of one- and two-dimensional nanostructure would be very large or even divergent \cite{Aoki:2001th,Narayan:2002cl,Lepri:2003fc,Lepri:2005gg,Wang:2012wy}. On the other hand, from changes in phonon dispersion relations and from the growing impact of surface scattering one may argue that dimensionality reduction hampers thermal transport \cite{Balandin:1998ua}, by limiting phonon mean free paths below the characteristic size of the nanostructure, be it the diameter of a nanowire, or the thickness of a thin film or a membrane, which imposes the so called "Casimir limit" \cite{Casimir:1938tf}. These two scenarios stem from different standpoints but are not incompatible: only few materials are truly one or two-dimensional and can compare to ideal statistical physics models, and at the same time the application of Casimir limit to phonon scattering in nanostructures is oversimplified.  

Experiments on different systems show a complex reality in which dimensionality reduction may either boost or limit thermal transport depending on the systems and the configuration of the measurements. Very high, even possibly diverging, thermal conductivity, and extremely long phonon mean free paths were measured in graphene \cite{Balandin:2008fv,Ghosh:2010fq,Chen:2012fc,Balandin:2011gk,Xu:2014gy},and carbon nanotubes  \cite{Chang:2008cp,Yu:2005dd}, whereas a considerable suppression of thermal conductivity was observed in silicon nanowires \cite{Li:2003bo,Hochbaum:2008hl,Chen:2008ig}, thin films and membranes \cite{Asheghi:1997ke,Ju:1999uy,Liu:2005jo,Liu:2011dx,ChavezAngel:2014be}.
Nevertheless the measurements of thermal conductivity in nanostructures are very challenging, and yield large uncertainties, as they are very sensitive to the experimental conditions. For example the actual value of the thermal conductivity of suspended graphene is still debated and experimental estimates range from 2000 to 8000 Wm$^{-1}$K$^{-1}$.
In addition, even if techniques for a spectroscopic characterization of thermal transport have been recently developed \cite{Minnich:2011gr,Regner:2013gh}, so far they could not directly ascertain  the origin of the observed enhancement or suppression of thermal conductivity. 
In this scenario molecular simulations of nanostructures emerge as powerful tools to bridge the gap between simple models and experiments.
The main advantage of molecular modeling is that complexity and details can be gradually introduced into models, thus realizing "gedanken experiments" that allow one to ascertain the origin of complex physical phenomena.

In the next section the simulation approaches to heat transport in bulk and low-dimensional systems are briefly outlined. The application of such simulation methods to elucidate the effects of dimensionality on thermal transport in carbon and silicon nanostructures are reported in sections~\ref{sec:carbon} and \ref{sec:silicon}, respectively. Section~\ref{sec:conclusions} summarizes the main findings and suggests future perspectives.

\section{Simulation Tools}
\label{sec:tools}

Several atomistic simulation tools are available to investigate thermal transport in nanostructures and compute the thermal conductivity ($\kappa$) of materials.  Simulation methods can be sorted into two classes, namely Lattice Dynamics (LD) and Molecular Dynamics (MD).
It is often useful to combine distinct complementary approaches, as they involve different approximations and limitations, which make them suitable to probe different transport regimes in systems of various size and complexity.
Combining different  methods sheds light on a broader variety of aspects of nanoscale heat transport.
Hereafter the main-stream LD and MD approaches are briefly described.

\subsection{Anharmonic Lattice Dynamics}

In the harmonic approximation the normal modes of vibrations of a periodic system of $N$ particles are determined by diagonalizing the {\sl dynamical matrix} $\mathbf{D}$ for each momentum vector $\mathbf{q}$:
\begin{equation}
  \mathbf{D}(\mathbf{q})\mathbf{e}_\lambda(\mathbf{q})=\omega^2_\lambda(\mathbf{q})\mathbf{e}_\lambda(\mathbf{q}),
  \label{eig}
\end{equation}
which provides the frequencies $\omega_\lambda(\mathbf{q})$ and the normalized displacement vectors $\mathbf{e}_\lambda(\mathbf{q})$.  
$\mathbf{D}(\mathbf{q})$ is defined as:
\begin{equation}
  D_{ij}(\mathbf{q})=\frac{1}{\sqrt{m_i m_j}}\frac{\partial^2 \Phi}{\partial x_{i,\alpha} \partial x_{j,\beta}}\exp(i\mathbf{r} _{ij}\cdot\mathbf{q})
  \label{dynmat}
\end{equation}
where $\Phi$ is the potential energy of the system at equilibrium, $m_i$ is the mass of atom $i$, and $\mathbf{r_{ij}}$ the distances between pairs of atoms $i$ and $j$, and $\alpha$ and $\beta$ indicate the Cartesian components. In a system of $N$ particles $\mathbf{D}$ is then a $3N\times 3N$ matrix. In a first principles framework, the matrix elements of $\mathbf{D}$ can be computed either by finite differences or by density functional perturbation theory \cite{Baroni:2001tn}.  

Considering that quantized lattice vibrations can be treated as particles, their propagation in the diffusive regime can be described using the Boltzmann transport equation (BTE), in analogy with the diffusion of a gas:
\begin{equation}
 \frac{\partial n_\lambda}{\partial t} + \mathbf{v}_\lambda\cdot \mathbf{\nabla} n_\lambda = \left( \frac{dn_\lambda}{dt}\right)_{scattering},
\label{boltz}
\end{equation}
considering, however, that the phonons are spin-less quantum particles that obey to Bose-Einstein statistics.
The righthand side of (\ref{boltz}) accounts for the scattering processes that create or destroy phonons: namely, anharmonic scattering processes, isotopic, defect and boundary scattering. 
The non-equilibrium occupation function can be written as $n_\lambda = n^0_\lambda+\delta n$, and,
assuming small temperature gradients ($\nabla T$), one can linearize  (\ref{boltz}) by treating $\delta n$ perturbatively. In stationary conditions the first term of (\ref{boltz}) vanishes, and the linearized BTE is written as:
\begin{equation}
  \mathbf{v}_\lambda\cdot \nabla T  \frac{\partial n^0_\lambda}{\partial T} = \left( \frac{dn_\lambda}{dt}\right)_{scattering}.
\label{lBTE}
\end{equation}
The linearized BTE can be solved at different levels of accuracy and complexity.
The simplest approach is to calculate the life time of each phonon, assuming that the population of all the other modes is the one at equilibrium ($n^0_\lambda$) \cite{Callaway:1959uo}. 
The resulting expression for $\kappa$ is the sum over the contribution of all the phonon modes, integrated to convergence over the first Brillouin Zone of the system,  sampled with a grid of $N_\mathbf{q}$ q-points: 
\begin{equation}
  \kappa = \frac{1}{N_\mathbf{q}}\sum_{\lambda,\mathbf{q}} \kappa_\lambda (\mathbf{q}) =  \frac{1}{N_\mathbf{q}} \sum_{\lambda,\mathbf{q}} C_\lambda (\mathbf{q}) \mathrm{v}^2_\lambda (\mathbf{q}) \tau_\lambda (\mathbf{q}) 
  \label{bte-smrt}
\end{equation}
where $C_\lambda$ is the heat capacity per unit volume of each vibrational state, $\mathrm{v}_\lambda$ is the component of the group velocity in the direction of transport and $\tau_\lambda$ is the phonon lifetime.
Even if it is approximated, this is a very useful expression that allows one to resolve the contribution to $\kappa$ of each phonon branch at each frequency $\omega$. 
\begin{figure}[b]
\sidecaption
\includegraphics[scale=.32]{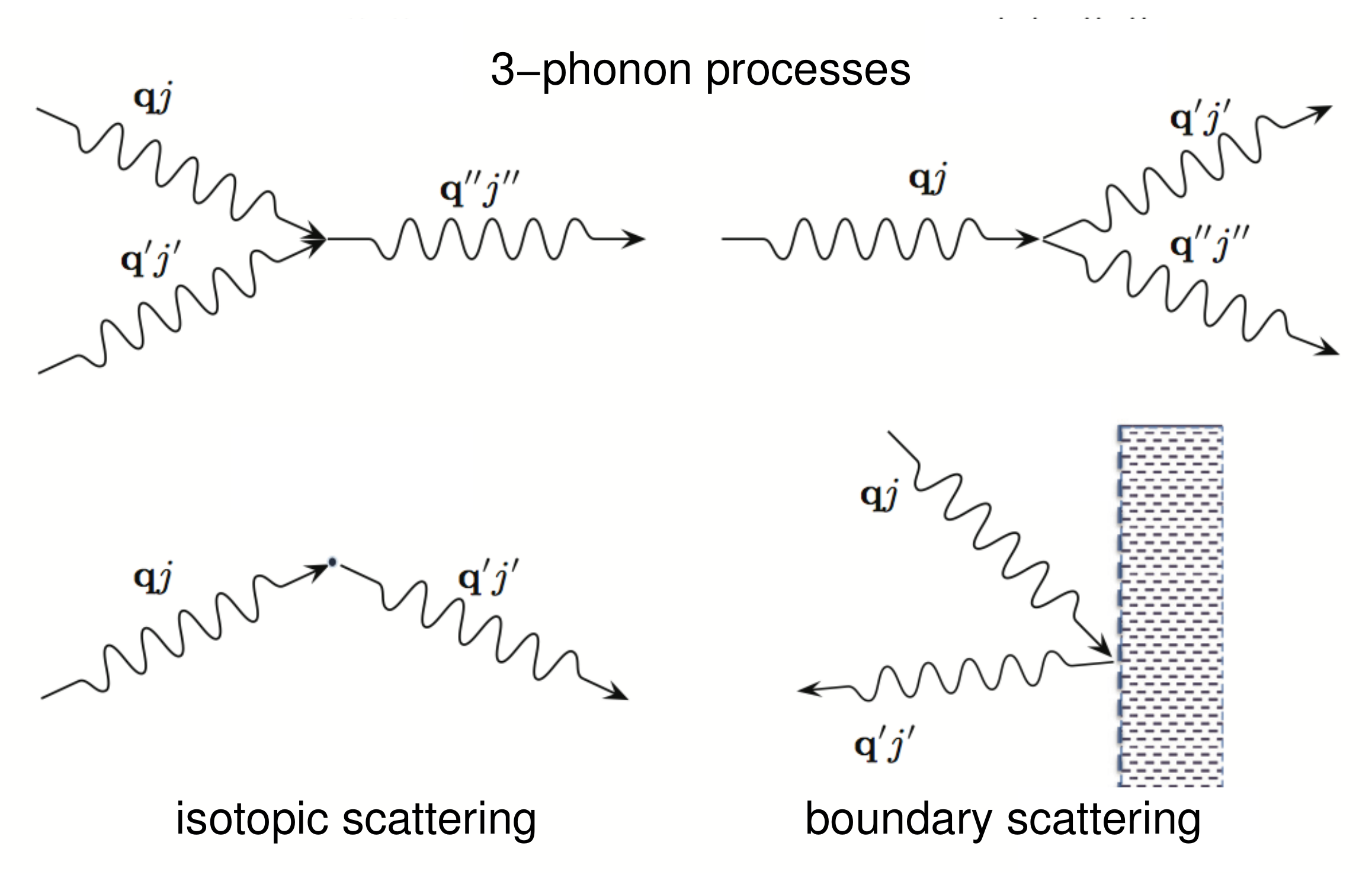}
%
%
\caption{Phonon scattering mechanisms in a crystalline thermal conductor (adapted from \cite{Fugallo:2013jl}). Copyright (2013) by The American Physical Society.}
\label{fig:diagrams}       
\end{figure}

Heat capacity and group velocities are usually obtained from the dispersion relations in harmonic approximation, though anharmonic corrections are also possible.  
To compute the phonon lifetimes $\tau_\lambda$, one has to consider all the scattering processes occurring in materials, namely phonon-phonon anharmonic scattering (normal and Umklapp), boundary and defects scattering (Figure~\ref{fig:diagrams}). 
In a perfectly crystalline material, neglecting electron-phonon interactions, the only possible phonon scattering mechanism is anharmonicity, and the main contribution comes from 3-phonon  processes. 
Two viable processes exist: either two phonons ($\omega_1$,$\omega_2$) annihilate into a third one ($\omega_3)$, or one phonon ($\omega_1$) decays into two phonons ($\omega_2$,$\omega_3$). Energy and momentum conservation determine the selection rules for three-phonon scattering: 
\begin{eqnarray}
 \omega_1(\bf{q}) \pm \omega_2(\bf{q}^\prime) - \omega_3(\bf{q}^{\prime \prime})  = 0 \\
 \bf{q} \pm \mathbf{q}^\prime -  \mathbf{q}^{\prime \prime} = \mathbf{Q}  
 \label{selection}
\end{eqnarray}
where $\mathbf{Q}$ is a reciprocal lattice vector.  Normal processes imply $\mathbf{Q}=0$, while in Umklapp processes $\mathbf{Q}$ is finite. Only Umklapp scattering processes dissipate energy thus contributing to limit $\kappa$. 
Computing  anharmonic scattering rates and phonon lifetimes ($\tau_{anh}$) requires the knowledge of the third derivatives of the interatomic potentials. 

In the single mode relaxation time approximation (SMRTA) phonon lifetimes are expressed as the inverse of spectral line width \cite{Fabian:1996wa,Maradudin1962}, treating on equal footing normal and Umklapp processes, which in fact contribute to scattering heat carriers with substantially different weights.   
In this approach the contribution from defect and boundary scattering, $\tau_{def}$ and $\tau_B$, add up to $\tau_{anh}$ through Matthiessen's rule:
\begin{equation}
  \frac{1}{\tau} = \frac{1}{\tau_{anh}} +\frac{1}{\tau_B} +\frac{1}{\tau_{def}} .
  \label{eq:Matthiessen}
\end{equation}  
While SMRTA has been widely used and can be implemented efficiently to treat large nanostructured systems \cite{Savic:2013tg}, it may result highly inaccurate at temperatures much lower than the Debye temperature, where normal scattering processes dominate, as it is the case for carbon based materials at room temperature, and especially for systems with reduced dimensionality.  \cite{Ward:2009iw,Fugallo:2013jl}. 
One can show that SMRTA provides a lower boundary to the thermal conductivity obtained by solving exactly the linearized BTE. 

Methods to solve exactly the linearized BTE have been proposed, either using a self-consistent iterative numerical approach \cite{Broido:2005kf}, or by minimizing a variational functional \cite{Fugallo:2013jl}. Exact approaches have demonstrated to be accurate and predictive for both bulk crystals  \cite{Broido:2007iu,Ward:2009iw} and nanostructures with low dimensionality \cite{Lindsay:2009cz,Lindsay:2014cg,Fugallo:2014bt,Cepellotti:2015ke}, especially when combined with {\sl ab initio} calculations of the harmonic and anharmonic force constants. 
So far the exact solution of the BTE has been applied only to crystalline systems with relatively small unit cells, also because the {\sl ab initio} calculations are computationally very demanding. In turn the variational BTE approach has been demonstrating to be extremely powerful to ascertain the effect of reduced dimensionality on heat transport in graphene and two-dimensional materials \cite{Fugallo:2014bt,Cepellotti:2015ke}.

\subsection{Equilibrium Molecular Dynamics}

Molecular Dynamics (MD) is a method designed to compute the  properties of solids or liquids by taking thermodynamic averages over a trajectory obtained integrating the classical equations of motion of the particles in the simulation box \cite{Allen1989a}. 
In the simplest case MD simulations are performed at equilibrium conditions in the microcanonical ensemble, i.e. at constant energy, volume and number of particles. 
As the typical size scale of MD simulations goes from few hundreds to millions of atoms, to represent extended (bulk) systems periodic boundary conditions are applied to the simulation cell.
Linear response and transport coefficients can be calculated from the fluctuations  of the respective conjugate flux via Green-Kubo relations~\cite{Kubo:1957we}.
In the case of heat transport, $\kappa$ is calculated from fluctuations of the heat flux via the heat flux autocorrelation function (HFACF).
For a system in equilibrium, in the absence of a temperature gradient, the net heat current averages to zero over time, but the integral of its correlation function is finite and proportional to its thermal conductivity.
The Green-Kubo expression for each component of the thermal conductivity tensor can be written as
\begin{equation}
\kappa_{\alpha \beta} =  \frac{1}{k_BT^2}\lim_{t\to\infty} \lim_{V\to\infty} \frac{1}{V} \int_0^t \langle J_{\alpha}(t') J_{\beta}(0) \rangle dt',
\label{eq:greenkubo}
\end{equation}
where $\mathbf{J}$ is the heat flux, $k_B$ is Boltzmann's constant, $T$ is the system temperature and $V$ its volume.
In systems with anharmonic interactions the HFACF should decay to zero for large $t$, and its integral should saturate at a constant value. In practice, at long times the HFACF becomes noisy due to poor statistical sampling, and the integral drifts or presents large oscillations because of this statistical noise. 
Therefore, $\kappa_{\alpha \beta}$ is taken as the stationary value of  (\ref{eq:greenkubo}) before it starts drifting due to the accumulated numerical noise.

While $\kappa$ is a second-order 3x3 tensor, one is normally interested in the diagonal components.
In the case of 1D materials, such as nanotubes and nanowires, the system can be oriented in such a way that only one component is not zero.
In  2D systems there are only two independent  non-vanishing components $\kappa_{xx},\kappa_{yy}$, which may be equivalent by symmetry, for example in the case of graphene.
For 3D materials with cubic symmetry, like bulk silicon or diamond, there are three equivalent non-zero components $\kappa_{xx}=\kappa_{yy}=\kappa_{zz}$.

The advantage of equilibrium MD is that one does not have to make any assumptions on the type of phonon scattering and all orders of anharmonicity are taken into account. In addition with empirical potentials one can simulate relatively large systems, up to 10$^7$ particles, with no specific requirements of being crystalline: amorphous, polycrystalline, defective and liquid systems can be studied. On the other hand, convergence of the numerical integration of Eq.~\ref{eq:greenkubo} in time and size needs to be checked thoroughly, and, depending of the material under investigation, it may occur only for very large samples and for fairly long time-scales. In addition, MD is based on Newtonian dynamics and quantum effects are not taken into account, thus it cannot provide quantitative predictions on $\kappa$ at temperatures much lower than the Debye temperature $(\Theta_D)$.

\subsection{Non-equilibrium Molecular Dynamics}

Another, perhaps more intuitive way of using MD is to simulate stationary non-equilibrium conditions. One defines two regions of the simulation cell as heat source and heat sink, and generates an energy flux between them, through a part of the system where the atoms dynamics is unperturbed. Following Fourier's law, at stationary conditions a temperature gradient $\nabla T$, proportional to the energy flux $J$, is established, and the thermal conductivity is given by the proportionality constant:
\begin{equation}
 \kappa=\frac{J}{\nabla T}.
\label{Eq:Fourier}
\end{equation}
\begin{figure}[htb]
   \sidecaption
   \includegraphics[scale=.44]{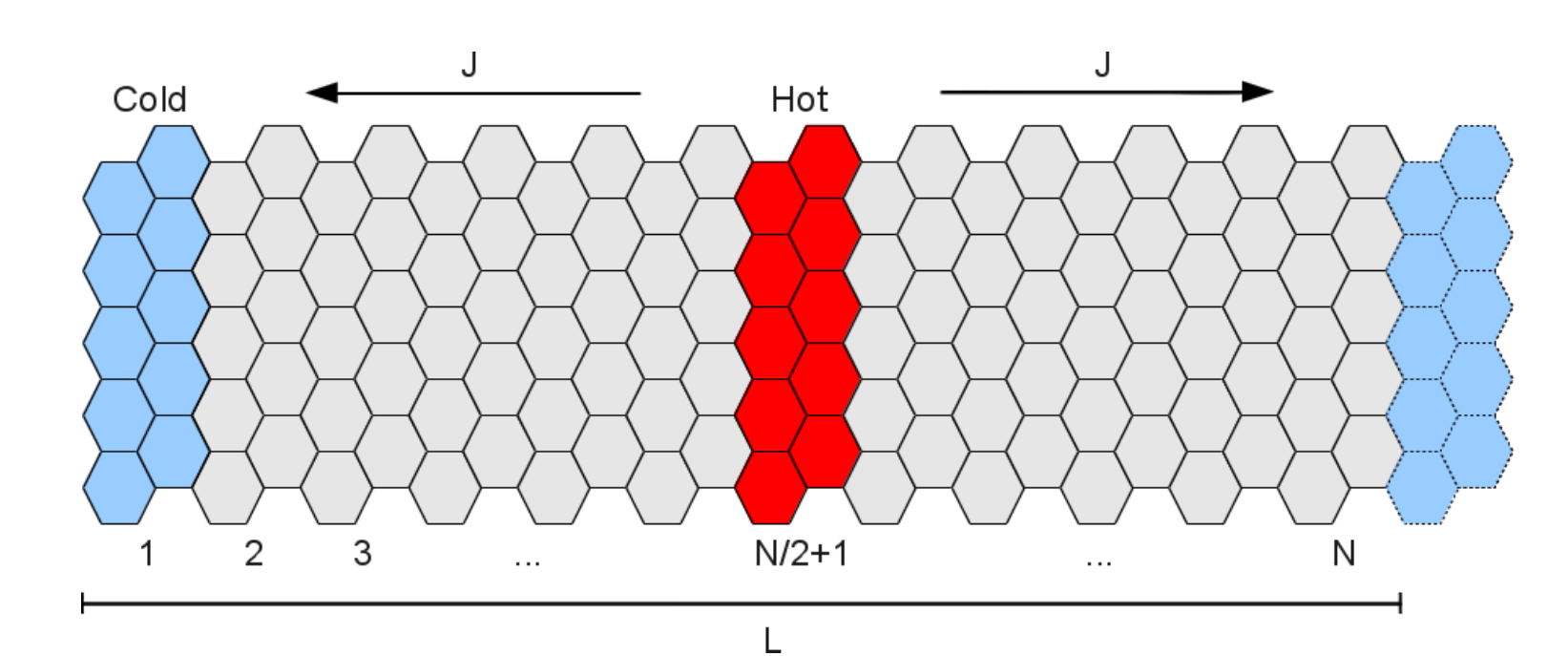} 
   \includegraphics[scale=.33]{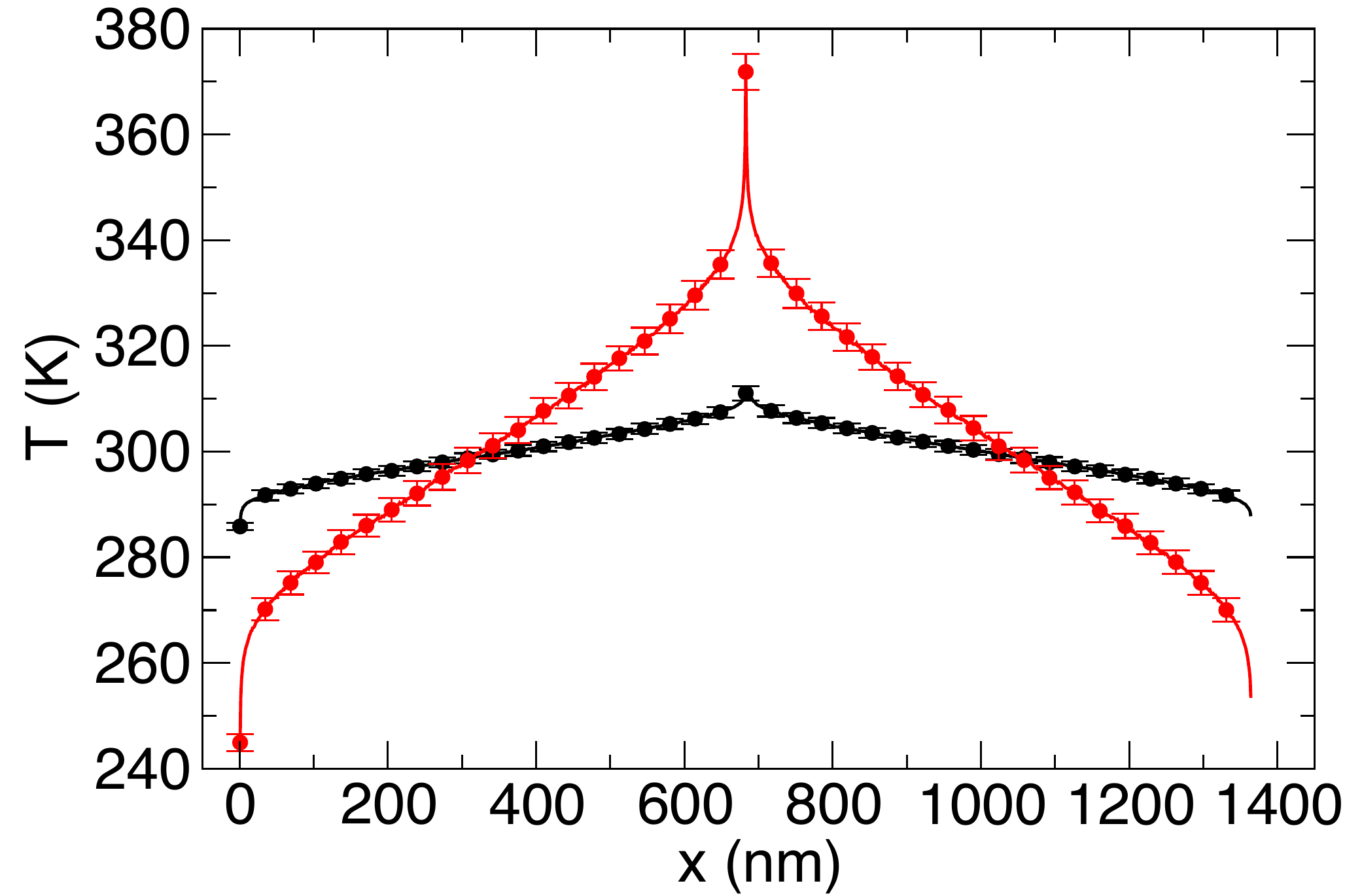} 
   \caption{Setup of a non-equilibrium molecular dynamics simulation (top panel) and temperature profile for a graphene patch 1.4 $\mu$m long with different imposed heat flux at stationary conditions (bottom panel).}
   \label{fig:NEMD}       
\end{figure}
A typical setup for NEMD simulations, is shown in Figure~\ref{fig:NEMD}. 
There are several ways of controlling the temperature in the thermal reservoirs. An option is to apply thermostats: stochastic local thermostats, e.g. Langevin, are preferable, as one would prefer to have fully phonon-absorbing reservoirs, to avoid that phonons travel through the heat sink uncontrollably if periodic boundary conditions are applied \cite{Jund:1999tb}. Alternatively one can apply the reverse-NEMD approach, in which the stationary heat flux is induced by swapping the momenta of particles between the heat source and sink. This is a particularly elegant method, since, together with a stationary heat flux, one automatically achieves energy and momentum conservation \cite{MullerPlathe:1997ub}.

NEMD simulations are closer in realization to experimental measurements, and are suitable to probe the onset of non-linear effects due to large temperature gradients. In turn, size scaling of NEMD to obtain the thermal conductivity of extended systems is tricky, as phonon mean free paths are truncated at the distance $l$ between the heat source and sink. One needs to perform simulations for a set of values of $l$ and extrapolate for $l$ going to infinity, as convergence can rarely be achieved. The standard way of extrapolating assumes that the inverse of $\kappa (l)$ is linear with $1/l$:\begin{equation}
 \frac{1}{\kappa (l)}= \frac{1}{\kappa_{\infty}} + \frac{C}{l}
\label{Eq:sizeNEMD}
\end{equation}
so that the intercept of the linear fit gives $1/\kappa_{\infty}$. In fact this extrapolation scheme assumes that the distribution of phonons that mostly contribute to heat transport have a narrow distribution of mean free paths, which is often not the case~\cite{Sellan:2010fl}. 
This feature of NEMD can also be exploited to resolve the relative contribution of phonons with a certain mean free path to the total $\kappa$ of a material, in accordance with thermal conductivity spectroscopy experiments \cite{Minnich:2011gr}.

If size scaling is performed correctly NEMD and equilibrium MD should provide results in agreement within statistical uncertainty, as it was demonstrate for several bulk systems \cite{Schelling:2002jl,He:2011wq,He:2012tq}.
More rigorously, it was demonstrated by renormalization group analysis and mode coupling theory that for momentum conserving systems there is a strict correspondence between the decay of the heat flux autocorrelation function, $C_{jj}(t) \propto t^\beta$, in Eq.~\ref{eq:greenkubo} and the function $\kappa(l)$ in NEMD \cite{Lepri:2003fc,Dhar:2008ij}. Although consensus has not been reached on the exact values of $\beta$ it is now well accepted that in three dimensions that $\beta>1$, in two dimensions $\beta=1$ and in one dimension $\beta<1$. These exponents imply that in one-dimensional systems $\kappa(l)$ diverges like $l^\alpha$ with $\alpha = 1-\beta$, in 2D $\kappa(l)\propto log(l)$, and in 3D $\kappa$ is finite \cite{Narayan:2002cl,Lippi:2000tj,Wang:2012wy,Saito:2010jo}. 
This picture, resulting from the combination of analytical calculations and numerical simulations on lattice models, has  stimulated further investigations on real systems but has not been confirmed either by experiments or by simulations of  nanostructures.
 
\subsection{Empirical interatomic potentials}

On crystalline systems with few atoms in the unit cell it is possible to perform {\sl ab initio} anharmonic lattice dynamics calculations using density functional perturbation theory and the $n+1$ theorem~\cite{Baroni:2001tn,Broido:2007iu,Fugallo:2013jl}.
On the other hand, given the large size and time scales required for the convergence of MD simulations, it is usually impossible or impractical to combine MD and first-principles approaches to compute thermal transport in nanostructures. For this reasons one has to rely on empirical potentials, whose accuracy and transferability need to be probed. 

For the materials considered in this book Chapter, mostly Carbon and Silicon, a few analytical empirical models have been shown to provide reliable estimates of the harmonic and anharmonic vibrational properties and of the total thermal conductivity of the most common bulk crystalline polymorphs. 
The most used many-body potentials for MD simulations of carbon nanostructures, such as graphene and carbon nanotubes, assume the analytical forms proposed by either Tersoff \cite{Tersoff:1989tr} or Brenner, suitably re--parametrized to reproduce correctly the phonon dispersion relations of graphene \cite{Lindsay:2010kq}. 
The Tersoff potential provides a sufficiently reliable estimate of the thermal conductivity of silicon and other semiconductors \cite{He:2012tq}. Alternatively the Stillinger-Weber potential has been extensively used to study thermal transport in silicon-based systems \cite{Schelling:2002jl}, and the environment dependent interatomic potential (EDIP) have been shown to provide an estimate of the thermal conductivity of bulk silicon very close to the experimental one \cite{Sun:2006cg}.


\section{Carbon based Nanostructures}
\label{sec:carbon}

\begin{figure}[htb]
   \sidecaption
   \includegraphics[scale=.67]{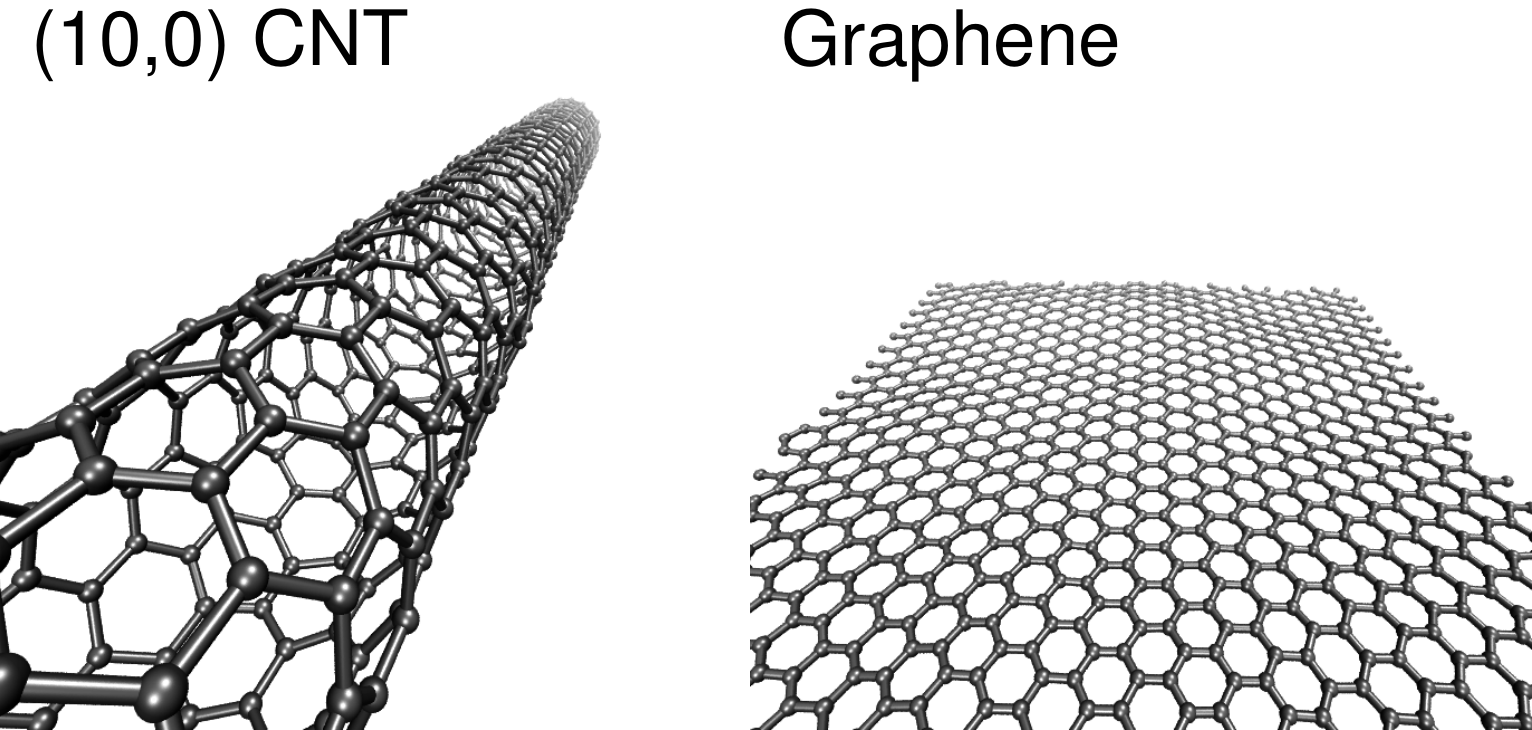} 
   \caption{Atomistic models of a (10,0) single wall carbon nanotube and of graphene.}
   \label{fig:carbonmodels}       
\end{figure}

The rich variety of nanostructures that can be formed from graphitic carbon provides an ideal platform to probe the effect of dimensionality on lattice thermal transport, and hopefully compare directly to non-linear models. 
Thermal transport in graphene and carbon nanotubes has been studied extensively searching for cases breakdown of Fourier's law and divergence of thermal conductivity with size \cite{Balandin:2011gk,Xu:2014gy,Chang:2008cp}.
While experimental measurements highlighted extremely high thermal conductivity and phonon mean free paths exceeding tens of nanometers, they could not provide a final assessment on the divergence of $\kappa$ in graphene and CNTs, due to the limited size of the samples, the presence of unavoidable intrinsic defects, and technical issues, such as, for example,  contact resistance. 
Atomistic simulations, using a combination of the techniques discussed in the previous section, are thus necessary to rationalize the experimental results.
In the following subsection we discuss heat transport in graphenen and in a (10,0) CNT, which is however representative of the general case of single wall CNTs. Atomistic models of these systems are shown in Figure~\ref{fig:carbonmodels}. 

\subsection{Graphene}

Being a single sheet of atoms, atomically flat, graphene is the only truly two-dimensional material that may be compared directly to non-linear two-dimensional models. However, graphene atoms can vibrate perpendicularly to the plane and form ripples. At finite temperature ripples confer stability to graphene impacting its thermal properties: for example negative thermal expansion coefficients are a consequence of out-of-plane vibrations and ripples.

\begin{figure}[htb]
   \sidecaption
   \includegraphics[scale=.67]{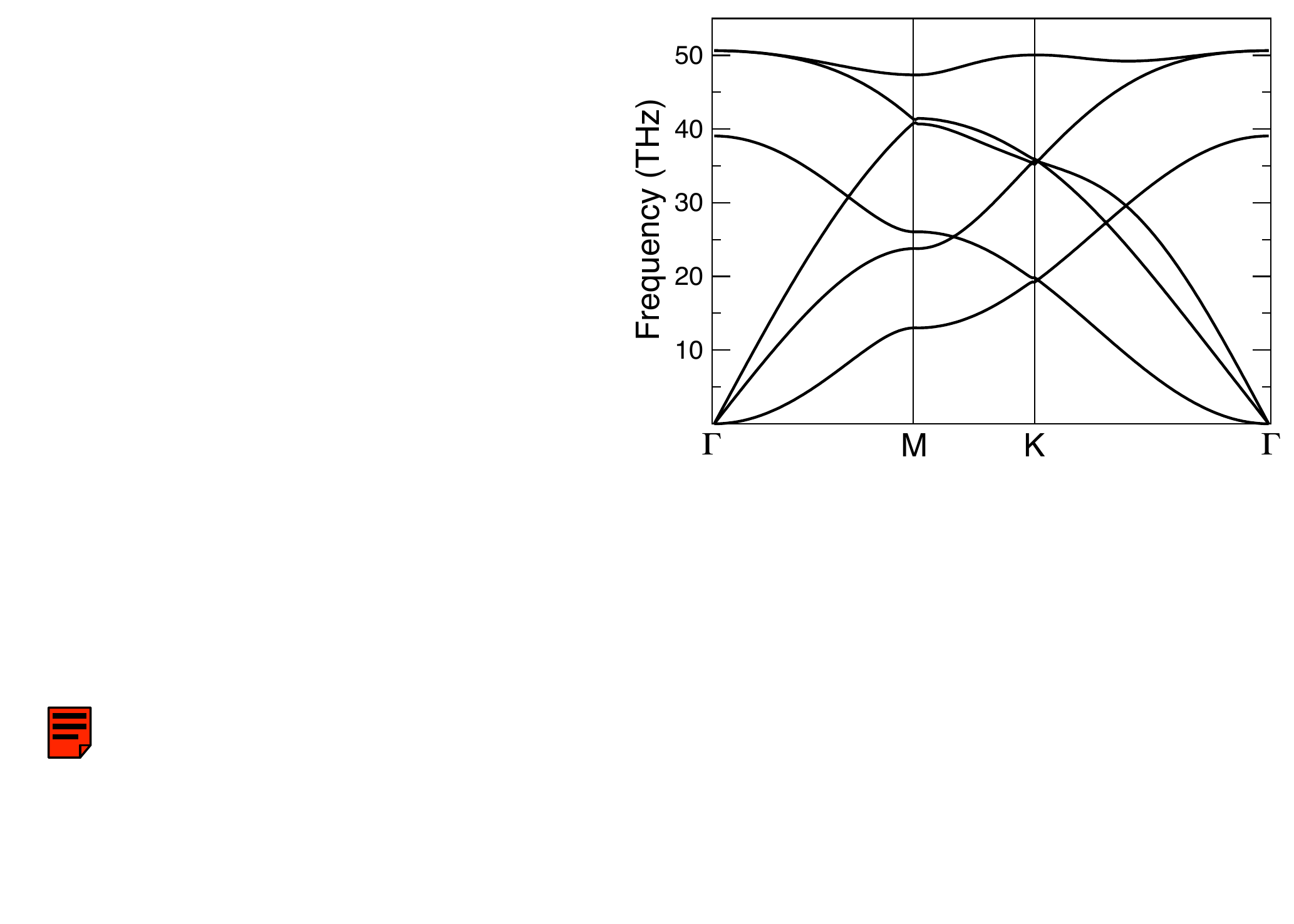} 
   \caption{Phonon bands of graphene computed with an optimized parameterization of the Tersoff potential \cite{Lindsay:2010kq}.}
   \label{fig:graphspectrum}       
\end{figure}

The phonon spectrum of graphene (Figure~\ref{fig:graphspectrum}) resembles very closely the one of graphite \cite{Nicklow1972}. The low frequency spectrum is characterized by a longitudinal in-plane mode with very high group velocity, a stiff in-plane transverse acoustic mode, and a softer flexural mode with quadratic dispersion near the Gamma point ($\omega\propto q^2$). The acoustic modes extend to frequencies as high as 42 THz, and are complemented at higher frequency by three optical modes that give rise to the extensively studied Raman peaks of sp$^2$ carbon.  

The earliest successful model of heat transport in graphite \cite{KLEMENS:2002vq} was based on a Debye approach for a two-dimensional gas of phonons with lower cutoff frequency of 4 THz. 
Since inter-planar interactions are neglected the model applies seamlessly also to graphene. Klemens and Pedraza neglected also out-of-plane vibration with the argument that these modes have very low group velocity and would not carry a significant amount of energy.
This model provided reasonable estimates of $\kappa$ in good, yet fortuitous agreement with experiments. More recent calculations based on a self-consistent solution of the BTE indeed demonstrated that flexural out-of-plane modes provide a large contribution to the thermal conductivity of graphene, due to their relatively high density at low frequency and their very long mean free path \cite{Lindsay:2010bi}. 

\begin{figure}[htb]
   \sidecaption
   \includegraphics[scale=.4]{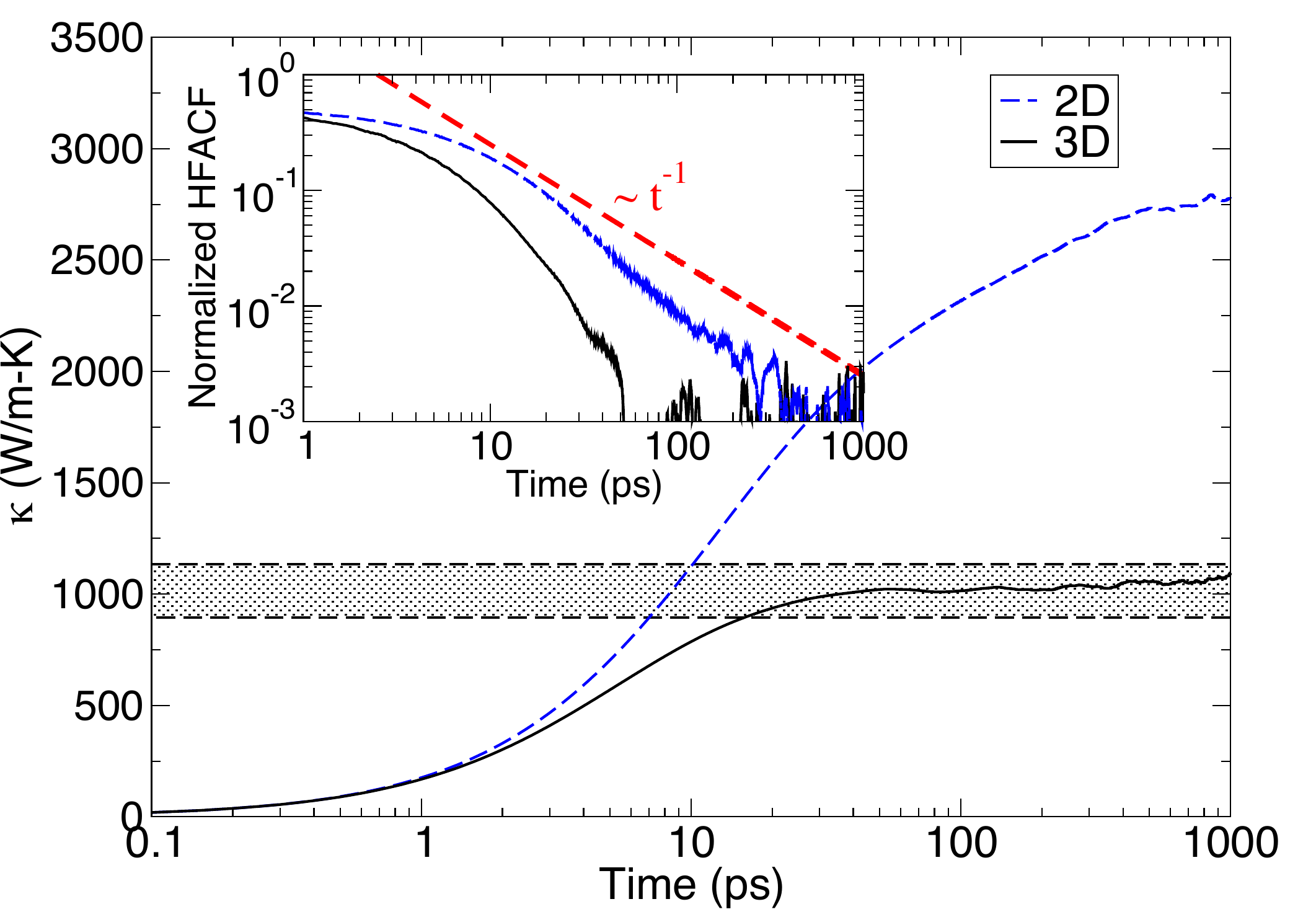} 
   \caption{Comparison of the heat flux autocorrelation functions (inset) and of their integrals (main graph) between a model of graphene with three-dimensional degrees of freedom (solid black lines)  and one, in which motion is confined in-plane (dashed blue lines), from \cite{Pereira:2013bo}. Copyright (2013) by The American Physical Society.}
   \label{fig:divergence2D}       
\end{figure}

Several studies addressed thermal transport in graphene by MD simulations providing estimates of $\kappa$ at room temperature from few hundreds to few thousands Wm$^{-1}$K$^{-1}$. Such variation of MD results stems from the sensitivity of $\kappa$ to the functional form and the parameterization of the empirical potentials and from the difficulties in converging MD simulations both in terms of time and size.
In addition any quantitative estimate of $\kappa$ of carbon-based materials at room temperature would suffer badly from the lack of quantum effects. Even if, given the very high Debye temperature of carbon-based materials, MD results can only be considered qualitative, equilibrium and non-equilibrium simulations provide useful insight in the physics of thermal transport in graphene. 
Equilibrium MD simulations show that $\kappa$ of graphene converges from above as a function of the size of the simulation cell, thus underlying the importance of sampling well low-frequency flexural modes, which assume a fundamental role in scattering heat carriers \cite{Pereira:2013bo}. 
A calculation of the heat flux autocorrelation function, well converged in terms of sampling, shows that for all the sizes of models considered it decays faster than the inverse of time, which makes the argument in  Eq.~\ref{eq:greenkubo} integrable, indicating that the system has a finite thermal conductivity.
Since the only conceptual difference between graphene models and two-dimensional non linear lattice models, for which $\kappa$ diverges logarithmically, is that in graphene the atoms can move out-of-plane, we performed an equilibrium MD simulation, in which the carbon atoms were frozen in-plane. 
This fully two-dimensional model exhibits $1/t$ decay of the heat flux autocorrelation function, indicating that diverging $\kappa$ is restored \cite{Pereira:2013bo} (Figure \ref{fig:divergence2D}).

\begin{figure}[htb]
   \sidecaption
   \includegraphics[scale=.32]{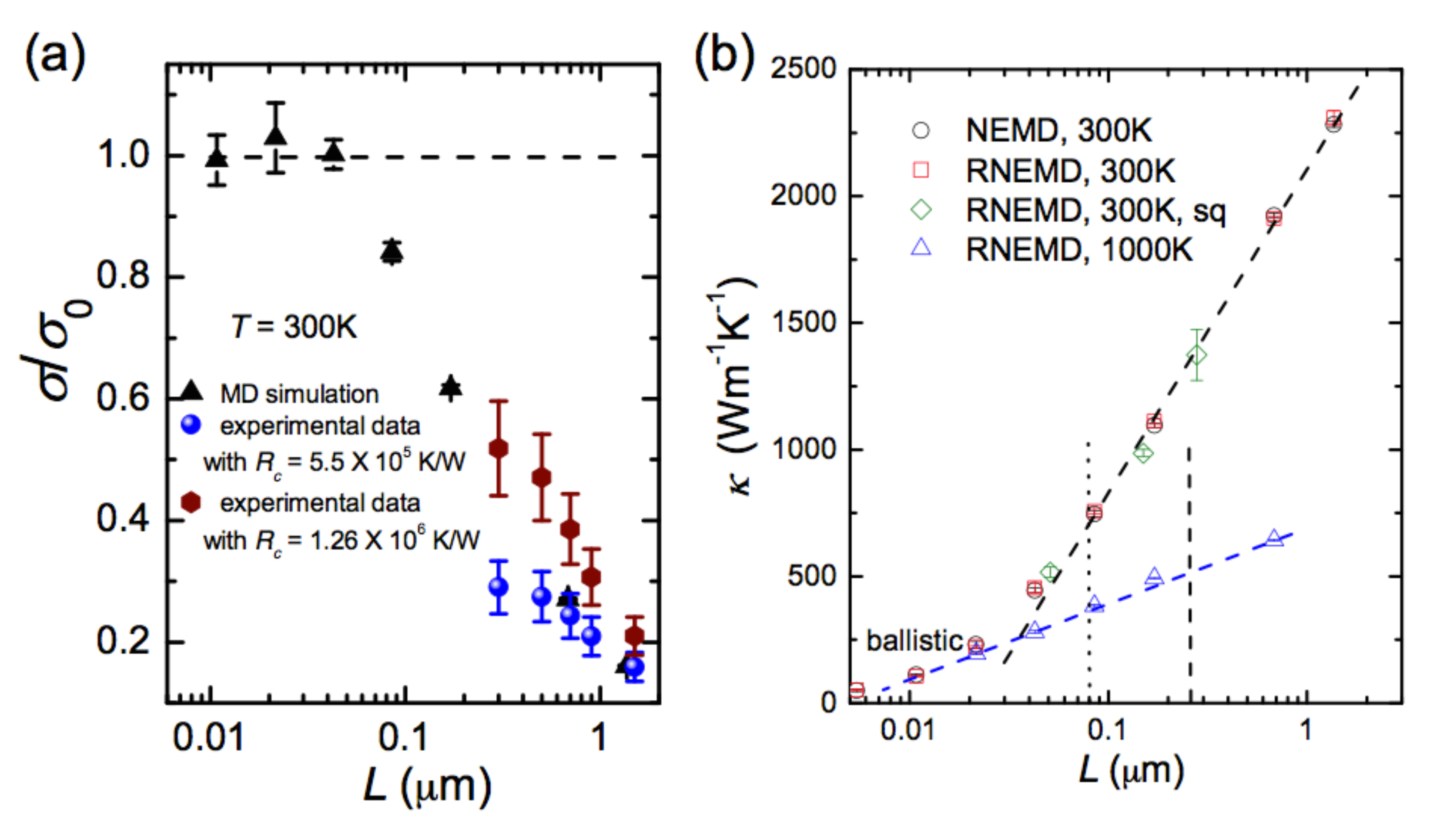} 
   \caption{(a) Normalized conductance of graphene as a function of length in non-equilibrium MD simulations and in experiments (a). Experimental data assuming two different contact resistances are reported. Transport is ballistic as long as $\sigma/\sigma_0 \sim 1$. (b) Thermal conductivity of graphene from non-equilibrium MD as a function of the size of the graphene model at T=300 and 1000 K. At both temperature $\kappa$ is not saturating for models 1.5 $\mu$m long. Different MD methods give consistent results. Adapted from \cite{Xu:2014gy}.}
   \label{fig:divergence}       
\end{figure}

While equilibrium MD simulations suggest that the thermal conductivity of graphene in the thermodynamic limit is finite, they do not provide direct information on the dependence of $\kappa$ on the size of the systems. 
The very large variation of the experimental estimate of $\kappa$, especially close to room temperature, may be partly justified by the different thermal conductivity as a function of the length of the graphene patches measured. 
Non-equilibrium MD simulations performed along with systematic experimental measurements showed that thermal transport in graphene at 300~K is ballistic up to $\sim 100$ nm (Figure ~\ref{fig:divergence}a). 
Simulations of larger models exhibit an apparently logarithmic divergence of the thermal conductivity both at room temperature and at 1000 K (Figure \ref{fig:divergence}b), in a agreement with measurements that show similar diverging trend up to 10 $\mu$m at room temperature \cite{Xu:2014gy}.
Non-equilibrium MD simulations of even larger systems, up to 100 $\mu$m, suggest that $\kappa$ saturates for the largest models considered \cite{Barbarino:2015kk}. 
Similarly, calculations performed solving the BTE self-consistently confirm the size dependence of $\kappa$ in the micrometer regime, but show that for larger lengths ($L$) $\kappa$ is not proportional to $\log(L)$ \cite{Lindsay:2014cg}. This study also underscored the substantial contribution of out-of-plane modes with long mean free paths to the total thermal conductivity: for example such contribution would amount to 75$\%$ for a 10 $\mu$m long graphene patch. 

\begin{figure}[htb]
   \sidecaption
   \includegraphics[scale=0.35]{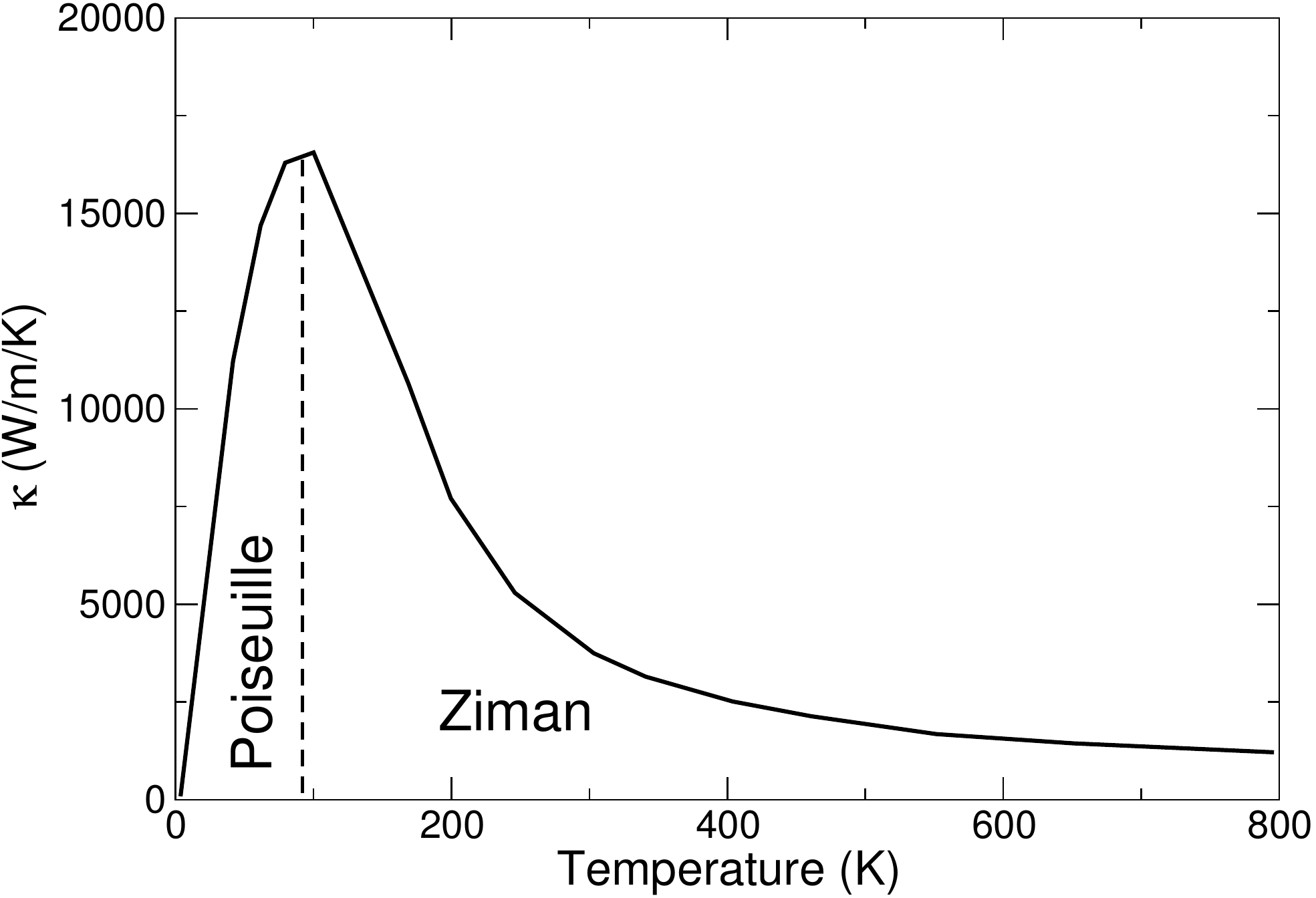} 
   \caption{Illustration of the different transport regimes in a graphene ribbon with a width of 100 $\mu$m. Data from \cite{Cepellotti:2015ke}.}
   \label{fig:hydrodynamic}       
\end{figure}

The application of a variational approach to BTE, which removes critical instabilities inherent in the self-consistent solution for two-dimensional systems \cite{Fugallo:2013jl}, finally resolved the debate on anomalous heat diffusion and the divergence of $\kappa$ in graphene. The exact solution of the BTE shows that the thermal conductivity of graphene converges in the thermodynamic limit \cite{Fugallo:2014bt}, in substantial agreement with equilibrium MD results, but convergence occurs for samples of the order of 1 mm. The reason is that the main heat carriers are not single phonons, but collective excitations with mean free paths of the order of 100 $\mu$m. 
This approach sheds new light on heat transport in two-dimensional systems, which is substantially different from three-dimensional materials. Due to the reduced dimensionality of the dual space, the selection rules for three-phonon scattering related to momentum conservation (Eq.~\ref{selection}) lead to a predominance of Normal processes over Umklapp processes. 
While in standard solids it occurs only at cryogenic temperatures, e.g. $\sim20$ K for Silicon, in layered materials this condition is verified over a broad temperature range, up to 800 K, leading to hydrodynamic phonon transport \cite{Cepellotti:2015ke,Lee:2015ex}. 
In graphene, as well as in other two-dimensional systems, such as graphane, hexagonal BN and MoS$_2$, the balance between non-dissipative processes and resistive ones dictates a Poiseuille regime at low temperature, before a peak of conductivity is reached, and a Ziman regime occurs at higher temperature (Figure~\ref{fig:hydrodynamic}). In both cases normal processes dominate. In Poiseuille flow normal scattering shifts the phonon distribution, and heat flux is eventually dissipated by extrinsic events, for example boundary scattering, whereas In the Ziman regime the shifted phonon distribution dissipates heat relaxing to equilibrium via Umklapp and isotopic scattering. This scenario has to be compared to that occurring in standard three-dimensional solids, in which transport regimes evolve from ballistic to diffusive in a narrow temperature window of few tens of K.   

\subsection{Carbon Nanotubes}

Momentum conserving one-dimensional models exhibit anomalous thermal conduction with length-dependent thermal conductivity $\kappa\propto L^\alpha$ \cite{Lepri:2003fc}. Carbon nanotubes, which can be grown up to tens of $\mu$m, were proposed as the most promising systems, for which anomalous heat conduction can be observed. Experiments agree that $\kappa$ is of the order of thousands Wm$^{-1}$K$^{-1}$, and is length dependent for lengths of the order of $\mu$m, thus supporting the hypothesis of a breakdown of Fourier's law \cite{Chang:2008cp,Balandin:2011gk}. 
However, in the light of the recent theoretical studies on graphene reported in the previous section, it is likely that experimental observations of anomalous heat conduction may be explained in terms of collective excitations and hydrodynamic transport regime. 
While this issue has not been clarified yet, it is worth analyzing the available results of lattice dynamics and molecular dynamics simulations, mostly focusing on single wall carbon nanotubes (SWCNT).

\begin{figure}[htb]
   \sidecaption
   \includegraphics[scale=0.45]{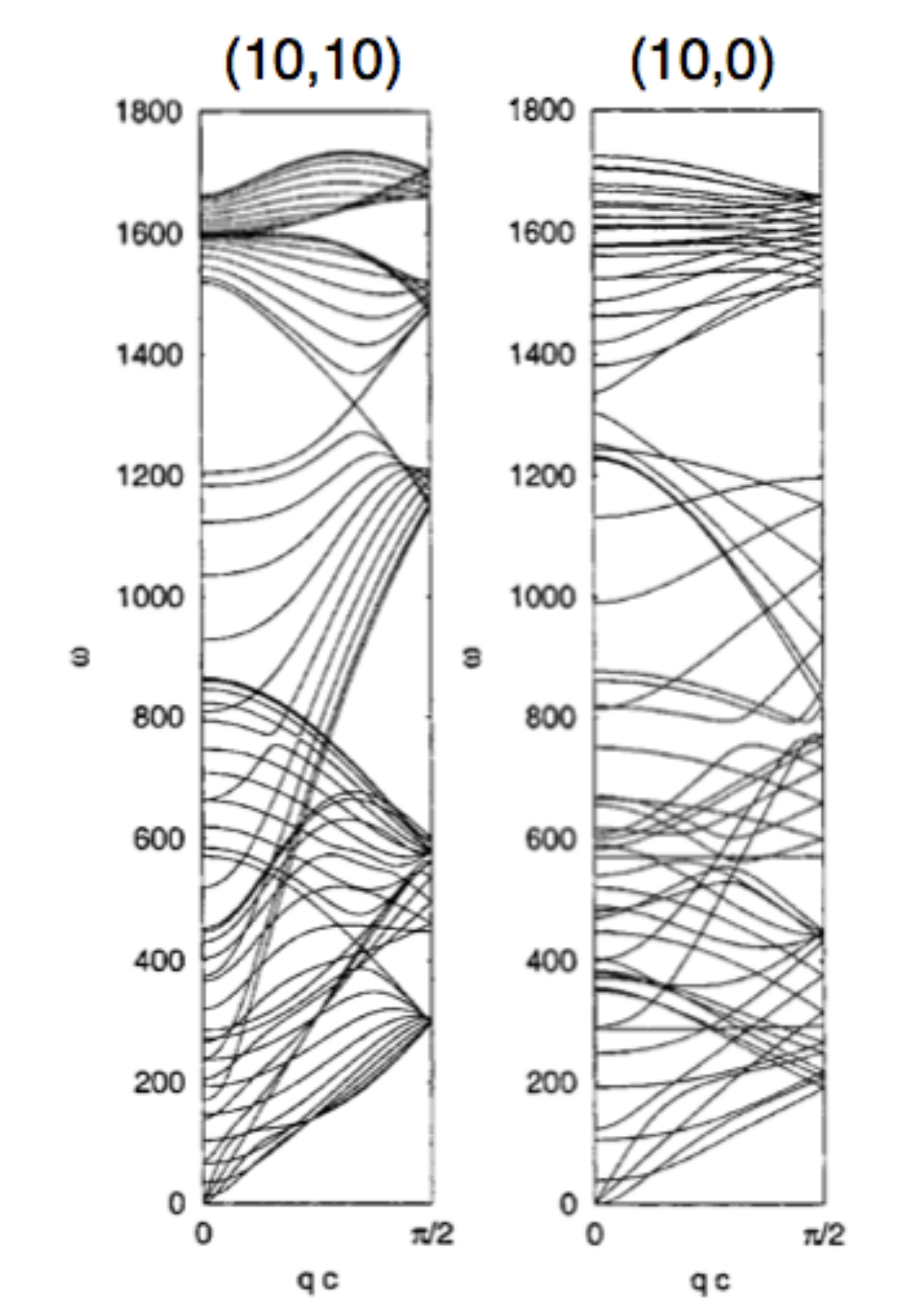} 
   \caption{Phonon dispersion relations of a (10,10) and of a (10,0) single wall carbon nanotube, adapted from \cite{Mahan:2004fd}. Copyright (2004) by The American Physical Society.}
   \label{fig:CNTphonons}       
\end{figure}

Ideal SWCNTs have cylindric symmetry and extend periodically in one dimension. Such geometry yields four invariances, namely three translations and one free rotation along the tube axis, which result in four acoustic modes. 
Two of them have quadratic dispersion ($\omega\propto q^2$) near the $\Gamma$ point and correspond to transverse sound waves. The longitudinal mode retains linear dispersion relation and has a group velocity higher than the torsional mode associated with the axial rotation, which has also $\omega\propto q$~\cite{Mahan:2004fd} (figure~\ref{fig:CNTphonons}). These features are independent on the chirality of the nanotubes, yet thinner wires have stiffer longitudinal modes and softer transverse flexure modes.
The large number of atoms in the unit cell gives rise to a large number of optical ``breathing-like" modes, which have nevertheless large group velocities at finite $q$-points. 

\begin{figure}[htb]
   \sidecaption
   \includegraphics[scale=0.25]{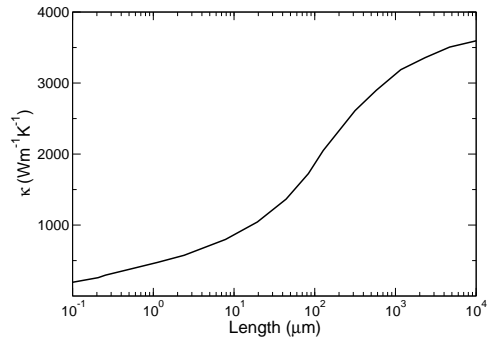} 
   \caption{Thermal conductivity of a (10,10) single wall carbon nanotube calculated by solving self-consistently the Boltzmann transport equation. Data from \cite{Lindsay:2009cz}.}
   \label{fig:CNTselectionrules}       
\end{figure}

The contributions from acoustic and  higher order breathing modes make the thermal conductivity of CNTs very high, as it was observed both in experiments and in simulations \cite{Kim:2001ei,Yu:2005dd,Berber:428480,Lukes:2007kx}. 
The reason for the extremely high thermal conductivity of CNTs is also related to their symmetry and one-dimensional periodicity. The selection rules for phonon-phonon interactions in a one-dimensional Brillouin zone limit the possibility of dissipative scattering (Umklapp), thus leading to very low scattering rates for flexural modes, and consequently to large $\kappa$ \cite{Lindsay:2009cz} (Figure~\ref{fig:CNTselectionrules}).

The possibility that heat transport would be anomalous \cite{Chang:2008cp}, implying divergence of $\kappa$ due to the low dimensionality of nanotubes, was explicitly addressed by MD simulations, which, however, provided contradicting results. Equilibrium MD simulations suggest a scenario similar to the one drawn for graphene, exhibiting integrable heat flux autocorrelation functions \cite{Donadio:2007ev,Pereira:2013dz}. On the other hand non-equilibrium simulations suggest anomalous heat conduction \cite{Savin:2009hr}. 
Such discrepancy may possibly arise from non linear effects related to the large temperature gradients that are necessarily used in non-equilibrium simulations. So high temperature gradients may also correspond to those induced in experiments, in which 100 K differences over few micrometers are not uncommon. Simulations show that in these condition thermal energy is transmitted efficiently by low-frequency mechanical waves that get coherently excited \cite{Zhang:2012ca}. In the light of the recent discovery of hydrodynamic transport in graphene \cite{Cepellotti:2015ke,Lee:2015ex}, it is also likely that similar phenomena happen in carbon nanotubes, thus leading to anomalous transport over the typical length scales adopted in simulations. However, this hypothesis is yet to be probed. 

A consequence of the low dimensionality of CNTs as well as of graphene is that thermal transport is extremely sensitive to any perturbation of the perfect crystallinity of these systems. MD simulations show that the thermal conductivity of carbon nanotubes is largely reduced by topological (Stone-Wales) and point defects. These defects affect the mean free paths of medium and high frequency phonons, while low frequency phonons are still transmitted ballistically over micrometer lengths. As a result $\kappa$ is much more sensitive to the concentration of defects rather than to their atomistic structure, and converges to similar values for different types of defects \cite{Sevik:2011ic}. $\kappa$ at room temperature can be reduced up to ten times by high concentrations of vacancies or di-vacancies. 
On the other hand the interaction either with other nanotubes, as in a CNT network, with a fluid medium or with a substrate dramatically affects the propagation of low frequency modes. Depending on the type of interaction with the substrate $\kappa$ can be also reduced about ten-fold \cite{Donadio:2007ev}, with respect to that of suspended CNTs. 
Extremely low $\kappa$ was predicted for CNT networks and pellets, however, in this case, heat transport is controlled by the very high contact resistance between two different CNTs, by the length of the CNTs, and eventually by the topology of the network \cite{Prasher:2009tx}.


\section{Nanostructured Silicon}
\label{sec:silicon}

Nanoscale silicon has been investigated thoroughly under several aspects, including thermal transport, due to its capital importance for technology, especially for applications in electronics and energy conversion. The constant reduction of the size of transistors, which has rapidly followed Moore's law down to few atomic layers, has made thermal dissipation at the nanoscale a crucial issue in nano-electronics. At the same time, finding a reliable and reproducible way of reducing the thermal conductivity of silicon without hampering its bulk electronic properties would enable silicon-based thermoelectric devices. Huge reduction of the thermal conductivity was observed for silicon nanostructures with reduced dimensionality, both for nanowires (1D)  and for ultra thin films and membranes (2D). The general understanding is that $\kappa$ is reduced by diffusive surface scattering, which becomes more and more effective the smaller the diameter of nanowires or the thickness of two-dimensional structures. 
However, this rather simplistic picture does not reconcile with the large enhancement of $\kappa$ predicted for non-linear models, and confirmed to a certain extent by experiments on carbon nanostructure.  

\begin{figure}[htb]
   \sidecaption
   \includegraphics[scale=.52]{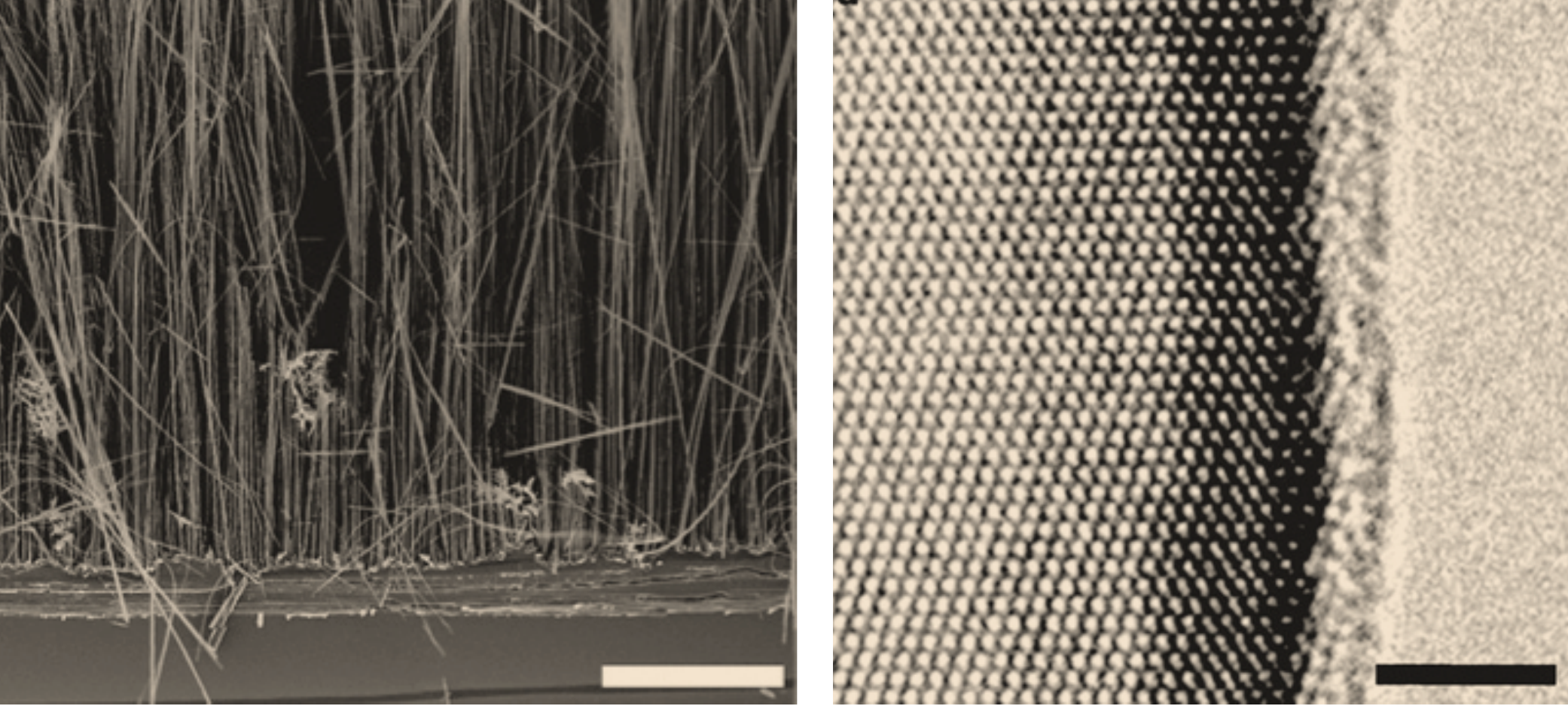} 
   \caption{SEM (left panel) and TEM (right panel) silicon nanowires produced by vapor-liquid-solid growth adapted from \cite{Hochbaum:2008hl}. The scale bar in the SEM image (left) corresponds to 10 $\mu m$ and the one in the TEM image (right) to 3~nm. The TEM image shows that the wire is crystalline but exhibits a disordered surface oxide layer about 4 atomic layers thick.}
   \label{fig:NW-SEMTEM}       
\end{figure}

Atomistic simulations can solve the conundrum, as they make it possible to disentangle different effects connected to nanostructuring. For example it is possible to check the effect of dimensionality reduction by simulating heat transport in ideally crystalline systems with diameter or thickness still unattainable to experiments.  In this respect it is worth noting that crystalline silicon nanowires with diameter as small as 1 nm were produced by oxide-assisted growth followed by etching \cite{Ma:2003ws} and membranes as thin as $\sim 8$ nm were obtained via advanced lithographic processes \cite{Shchepetov:2013gl}. However, their surface structure is subject to reconstruction, oxidation and roughness (see Figure~\ref{fig:NW-SEMTEM}), which cannot be easily controlled during  fabrication or simply upon exposure to standard environmental conditions, and measuring thermal transport in these systems is extremely challenging.  

\subsection{Silicon Nanowires}

Thermal transport in silicon nanowires was initially studied to probe whether  silicon nanostructures could be employed as high-performance thermoelectrics. Assuming that phonons would be scattered diffusively by surfaces, a significant reduction of phonon mean free path was expected, leading to highly reduced thermal conductivity.
Early predictions from simplified kinetic models, assuming that phonon mean free paths are chopped by diffuse surface scattering, were actually confirmed by the first equilibrium MD simulations, in which the interaction between silicon atoms was modeled using the Stillinger-Weber potential \cite{Volz:1999wq}. 
In this pioneering study the authors found that the thermal conductivity of model silicon wires with frozen boundary conditions, i.e. constrained surface atoms, is up to two orders of magnitude lower than the reference bulk value and is directly proportional to the diameter of the wires. The observation that $\kappa$ is temperature independent over a broad range of $T$, and that a model based on the BTE matches well the MD data lead to the conclusion that surface scattering is the key player in determining thermal transport in thin silicon nanowires. 
The measurements that followed, even though on wires with much larger cross section, confirmed the trends predicted by simulations \cite{Li:2003bo}.  
Nevertheless these early models do not exhibit realistic surface structuring, nor are the changes in phonon dispersion relations properly taken into account. 

\begin{figure}[htb]
   \sidecaption
   \includegraphics[scale=.54]{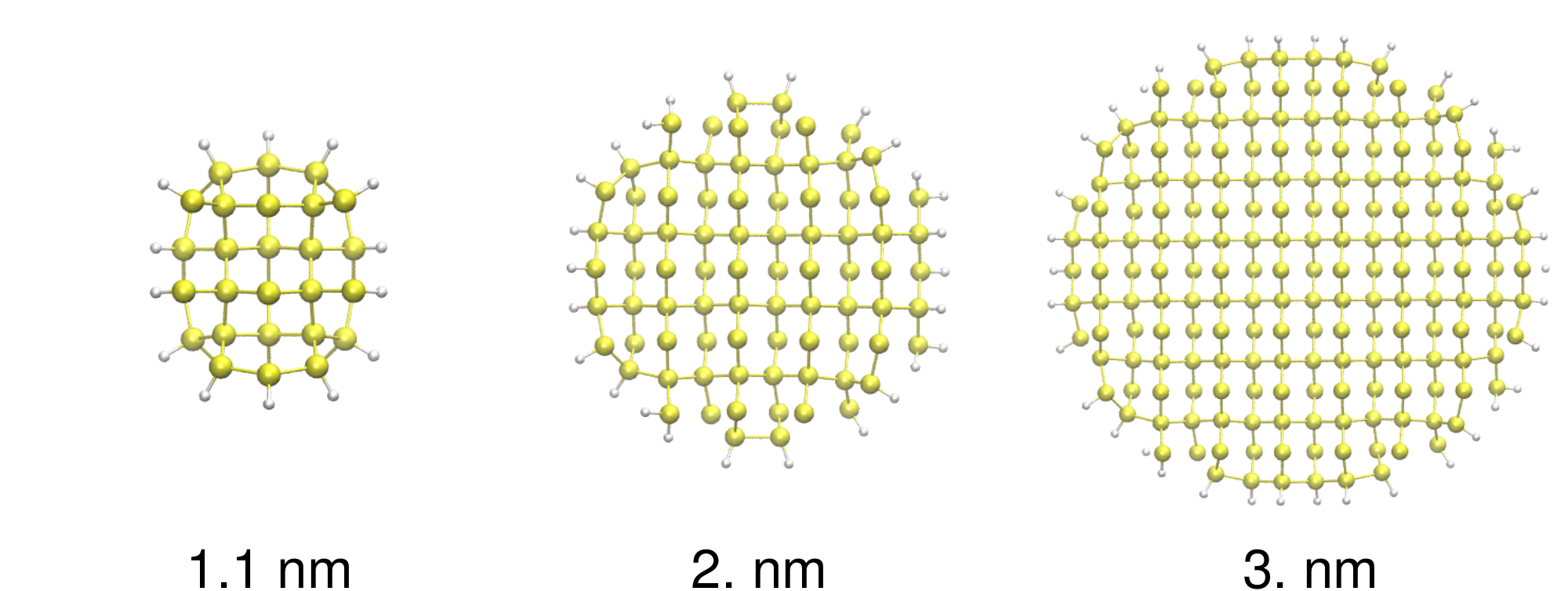} 
   \caption{Crystalline thin silicon nanowire models with hydrogenated  and ideally reconstructed surfaces  \cite{Vo:2006kd}.}
   \label{fig:SiNWmodels}       
\end{figure}

In examining the effect of dimensionality reduction in thin silicon nanowires, one should first consider how the phonon dispersion relations change with respect to the bulk reference for ideally crystalline wires, so that surface effects are ruled out completely. 
Let us consider for example three crystalline SiNWs with diameter of 1.1, 2 and 3 nm grown in (001) direction. The surface reconstruction of these wires, shown in Figure~\ref{fig:SiNWmodels}, was formerly optimized using density functional theory  \cite{Vo:2006kd}, and was proven to be stable also when the interaction between silicon atoms are modeled using the Tersoff bond-order interatomic potential.

\begin{figure}[htb]
   \sidecaption
   \includegraphics[scale=.45]{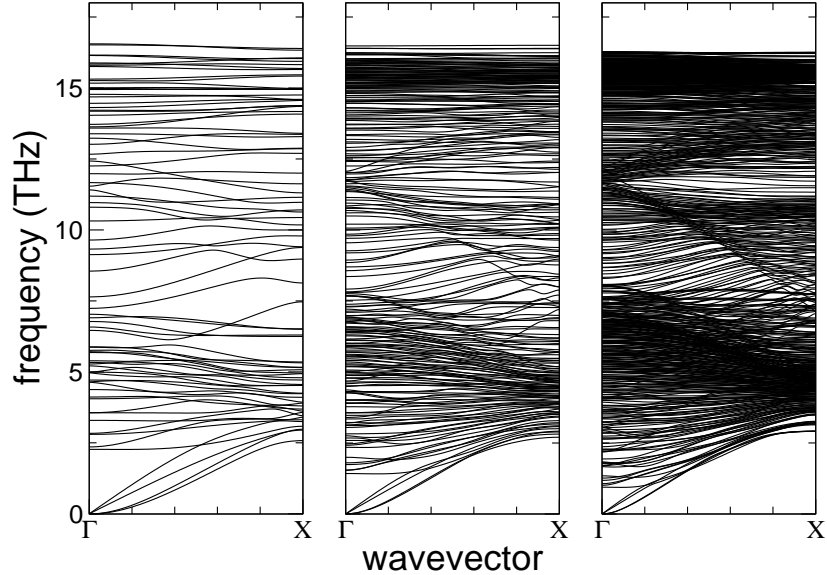} 
   \caption{Phonon dispersion relations of crystalline silicon nanowires with diameter of 1.1 nm (left panel), 2 nm (center) and 3 nm (right). High frequency Si-H bending and stretching modes are not shown.}
   \label{fig:phononsSiNW}       
\end{figure}

The phonon dispersion relations of these systems computed by harmonic lattice dynamics using Eq.~\ref{eig} are shown in Figure~\ref{fig:phononsSiNW}. 
As in the case of carbon nanotubes, the most immediate consequence of dimensionality reduction is that nanowires exhibit four acoustic modes, corresponding to three translational and one rotational invariant transformations. 
The two transverse acoustic modes, also dubbed flexural modes, have quadratic dispersion in the vicinity of the $\Gamma$ point. Calculations show that the thinner the wire, the softer these flexural modes are, and their dispersion relations remain quadratic for larger wavevectors. 
\begin{figure}[htb]
   \sidecaption
   \includegraphics[scale=.32]{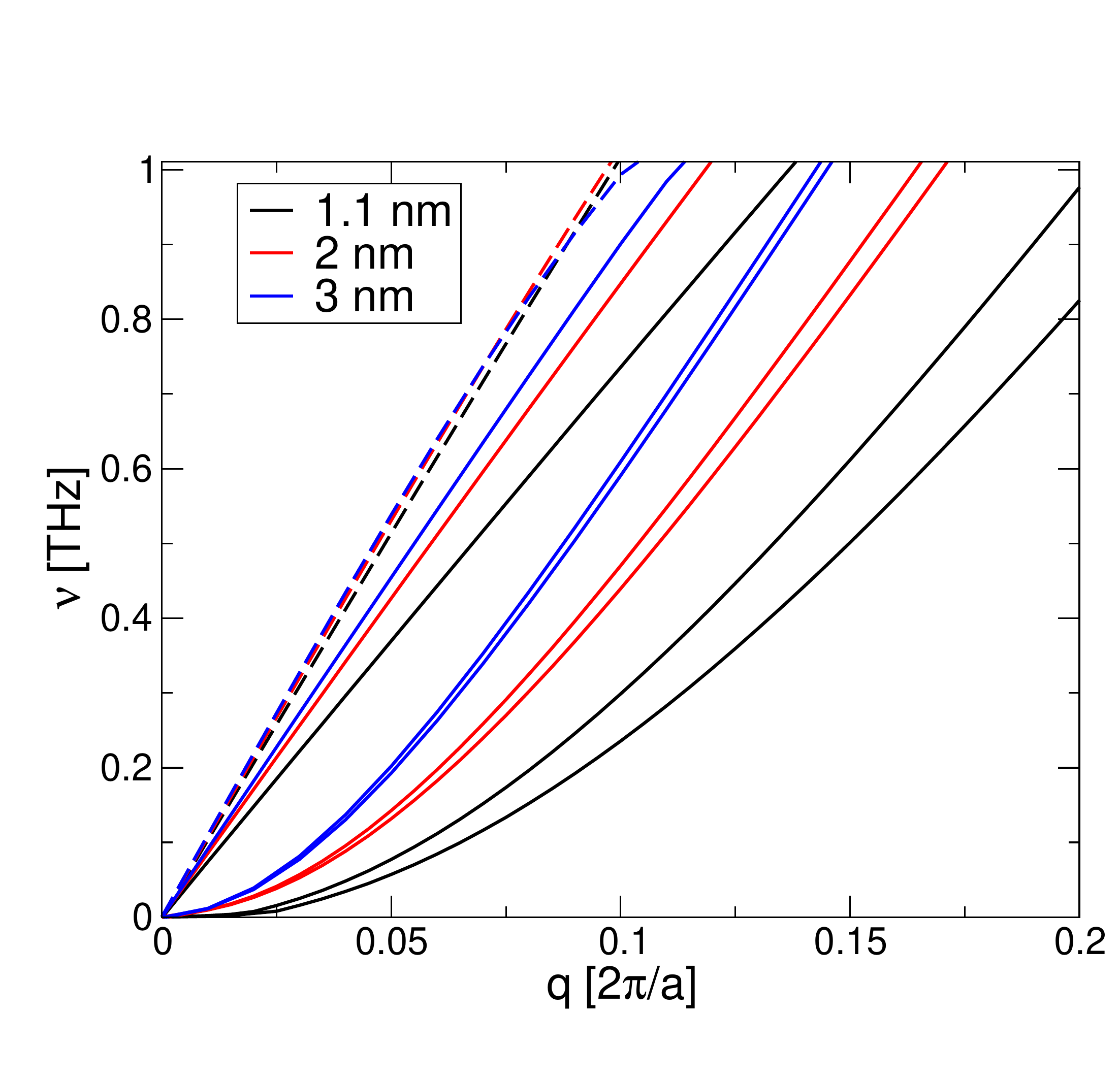} 
   \caption{Comparison of the dispersion relations of the acoustic phonons for thin silicon nanowires with different diameters. Flexural modes with $\omega \propto q^2$  and torsional modes with $\omega \propto q$ are plotted with solid lines, while longitudinal modes are plotted with dashed lines. }
   \label{fig:acousticSiNW}       
\end{figure}
The other two acoustic modes have linear dispersion in $q$ and correspond to torsional and longitudinal waves, respectively. 
Both the $q^2$ dispersion of the flexural modes and the presence of the torsional mode, originating from the cylindric symmetry, are a direct consequences of dimensional reduction and are independent on the material that constitutes the nanowires, the way interatomic forces are modeled or the surface structure. This scenario is very different from that of one-dimensional non linear models, which exhibit only a single longitudinal branch, since motion is restricted in one dimension. 

The longitudinal acoustic (LA) modes exhibit higher group velocities, which are not affected by the diameter of the wires, whereas the torsional modes become softer the thinner the wires (Figure~\ref{fig:acousticSiNW}). 
Due to the fairly large number of atoms in the unit cell of the nanowires considered here, higher frequency phonon branches appear at relatively low frequency. These modes provide viable channels to heat transport, however their group velocities remains limited, and significantly smaller than those of the acoustic phonons of bulk silicon.   
It is worth noting that the acoustic modes in silicon nanowires span a much more limited range of frequencies ($<5$ THz) than in bulk, where LA modes can extend up to 12 THz. 
From these observations it is difficult to draw precise conclusions on the thermal conductivity of crystalline wires, especially on its dependence on the diameter, but one would expect a reduction with respect to bulk, independently of surface scattering \cite{Balandin:1998ua}. 

\begin{figure}[htb]
   \sidecaption
   \includegraphics[scale=.27]{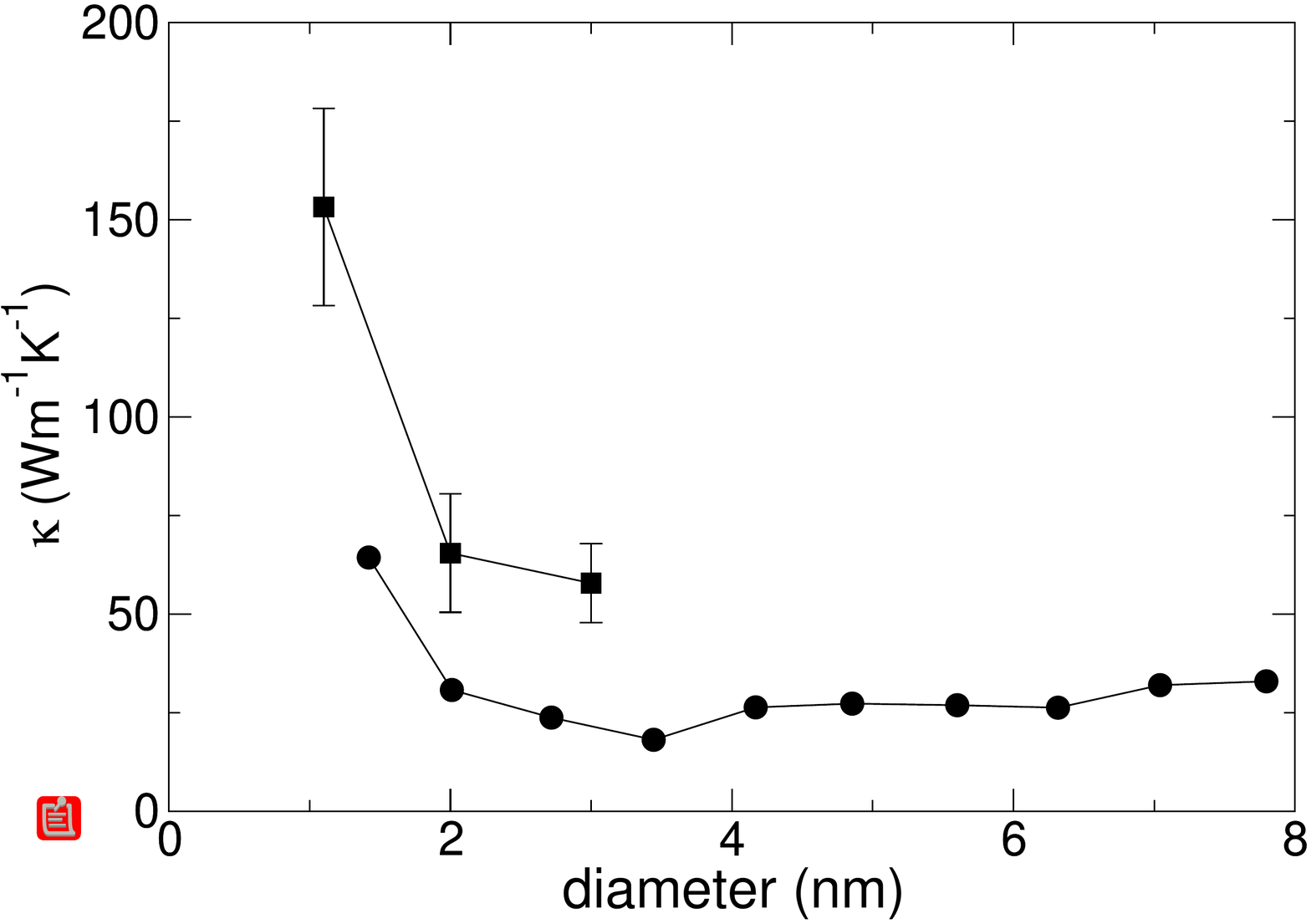} 
   \caption{Thermal conductivity of crystalline silicon nanowires at room temperature computed by non-equilibrium  \cite{Ponomareva:2007fj} and equilibrium molecular dynamics \cite{Donadio:2009fo}.}
   \label{fig:kappacrystalline}       
\end{figure}

Both equilibrium and non-equilibrium MD simulations demonstrate a non-monotonic dependence of $\kappa$ on the diameter of crystalline wires with similar structure \cite{Ponomareva:2007fj,Donadio:2009fo}.   
NEMD data from \cite{Ponomareva:2007fj} in Figure~\ref{fig:kappacrystalline} show that $\kappa$ decreases reducing the diameter, reaching a minimum for wires $\sim 3.5$ nm thick.  
Below this diameter $\kappa$ grows for thinner wires, and may even approach the bulk reference ($\sim 200$ Wm$^-1$K$^-1$), as in the case of the EMD estimate for the 1.1 nm wire. 
The location of the minimum of $\kappa$ would depend on the details of the model, including crystalline orientation of the wire, and surface reconstruction, as well as on the simulation setup, but it is a reproducible feature in nanowires. 
The increasing $\kappa$ of the thinnest silicon nanowires aligns to the general trend of enhanced thermal conductivity in systems with reduced dimensionality, observed also for carbon based materials. However, no simulation work has so far observed tendency to divergence, or enhancement of $\kappa$ beyond the bulk value. 

\begin{figure}[htb]
   \sidecaption
   \includegraphics[scale=.25]{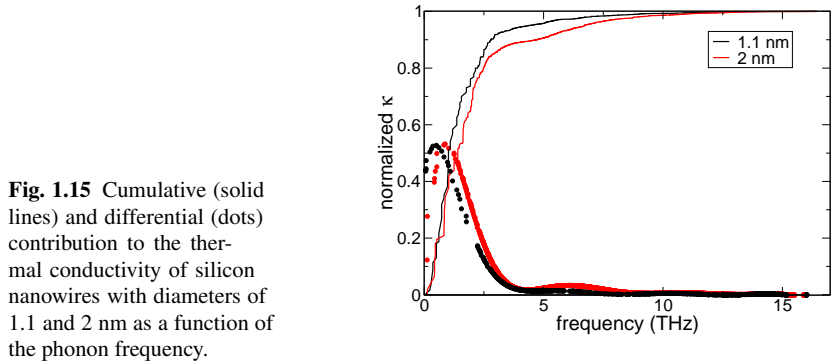} 
   \caption{Cumulative (solid lines) and differential (dots) contribution to the thermal conductivity of silicon nanowires with diameters of 1.1 and 2 nm as a function of the phonon frequency.}
   \label{fig:NWaccu}       
\end{figure}

An analysis of the spectral contribution to $\kappa$ as a function of phonon frequency in wires of 1.1 and 2 nm~\ref{fig:NWaccu}, obtained from Eq.~\ref{bte-smrt} in which lifetimes were computed by MD, shows that the distribution of heat carriers is shifted toward lower frequencies with respect to the bulk. This observation complies with the smaller frequency range of acoustic phonons in nanowires, and with the presence of flat bands with fairly small group velocities at higher frequency, which limits significantly phonon mean free paths. 
 This type of analysis yields a detailed explanation of the origin of the minimum in thermal conductivity, which takes into account the combination of density of vibrational states, group velocities and the phonon lifetimes.
Experimental verification of the prediction of a minimum of $\kappa$ as a function of diameter is nevertheless still lacking, since surface scattering would completely bleach the direct effect of dimensionality reduction.

\begin{figure}[htb]
   \sidecaption
   \includegraphics[scale=.32]{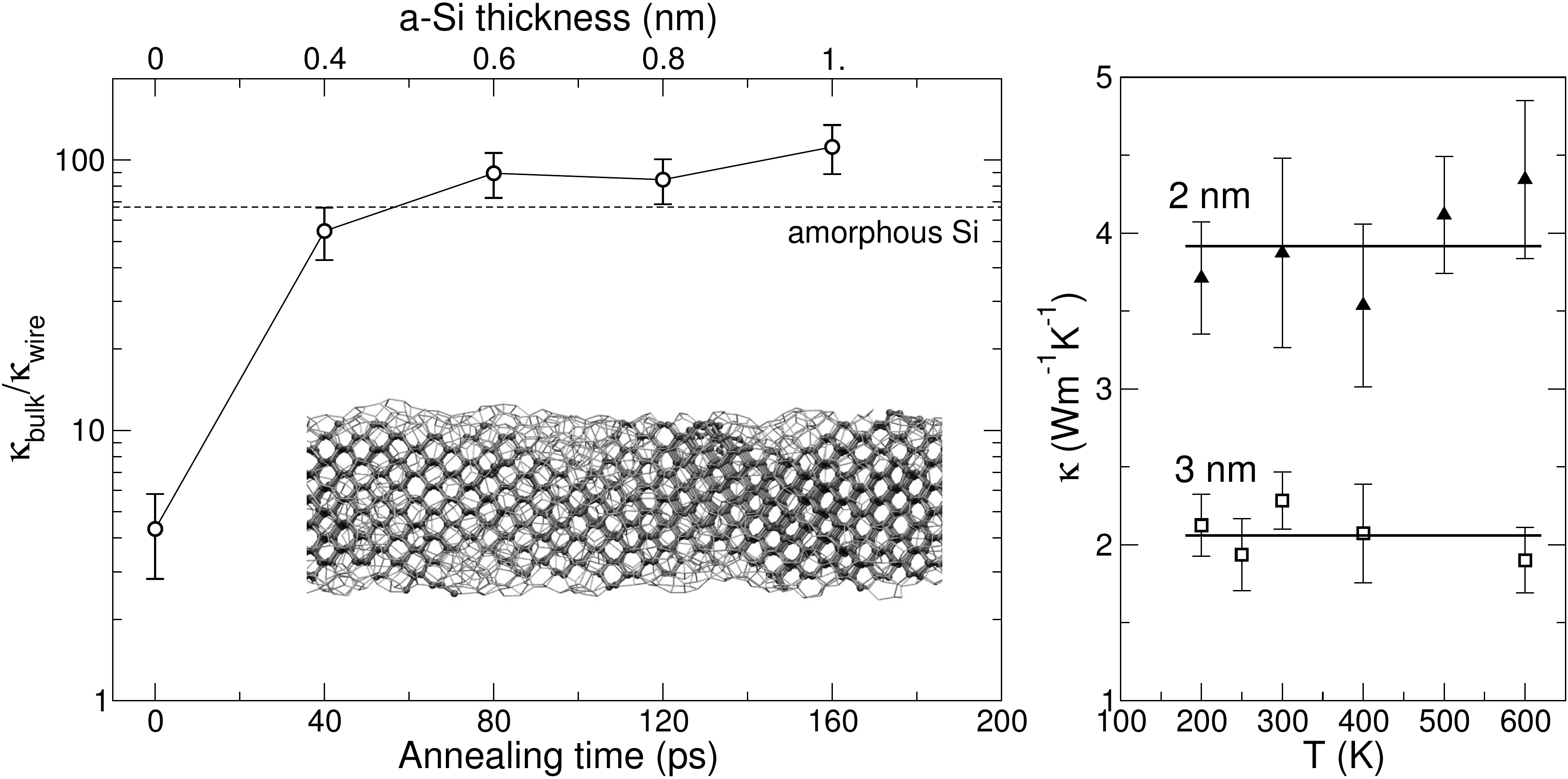} 
   \caption{(left panel) Thermal conductivity of a core-shell silicon nanowire with a diameter of 3 nm as a function of the thickness of amorphous shell from equilibrium molecular dynamics . The amorphous shell is generated by thermal annealing close to the melting temperature of silicon, and its thickness is proportional to annealing time. The dashed line corresponds to the thermal conductivity of bulk amorphous silicon computed by molecular dynamics using the same empirical potential~\cite{He:2011wq}. (right panel) thermal conductivity of core-shell nanowires with diameter of 2 and 3 nm as a function of temperature \cite{Donadio:2010kp}.}
   \label{fig:roughwire}       
\end{figure}
Thermal measurements on silicon nanowires with larger diameters have highlighted the role of  surface scattering of phonons in dictating a very low thermal conductance \cite{Chen:2008ig}. 
MD simulations predict that even more dramatic reduction of $\kappa$, up to 100 times lower than the bulk value at room temperature, can be achieved in the thin silicon wires in Figure~\ref{fig:SiNWmodels} covered with a thin layer of amorphous material \cite{Donadio:2009fo,Donadio:2010kp}. 
Given the very large surface to volume ratio, these systems are extremely sensitive to surface modification. Figure~\ref{fig:roughwire} shows that the reduction of thermal conductivity (plotted as the ratio of $\kappa_{bulk}/\kappa_{wire}$) is rather abrupt with the thickness of the amorphous shell: a rough layer of $\sim 6$ \AA\ of amorphous silicon, corresponding to about five atomic layers, is sufficient to abate $\kappa$ to 90 times less than bulk crystalline silicon at room temperature. 
Further surface amorphization does not induce further significant reduction of $\kappa$.
Remarkably, the thermal conductivity of these crystalline core/amorphous shell ultra-thin wires, for which the most of the volume remains anyway crystalline, can be lower than that of amorphous silicon computed with the same modeling techniques and the same empirical potential \cite{He:2011wq}.
These wires are among the few systems that have been predicted to break the {\it amorphous limit} of $\kappa$ at constant chemical composition \cite{Chiritescu:2007ee}.  
However, while several experiments have demonstrate a drastic reduction of $\kappa$ in nanowires, no measurements on ultra-thin wires with diameter of $\sim 3$ nm have been performed so far, thus theoretical predictions of $\kappa$ below the amorphous limit have not yet been confirmed. 

The temperature dependence of $\kappa$ (Figure~\ref{fig:roughwire}) and an analysis of the relative contributions of phonons to thermal transport demonstrates analogies to the characterization of heat carriers in disordered solids \cite{Allen:1993uc}.
As in amorphous silicon, vibrational modes in core-shell wires can be classified into low-frequency propagating phonons, which can be described by BTE, and higher-frequency diffusive modes, which contribute to heat transport via an energy transfer mechanism that can be described in harmonic approximation using Fermi's golden rule \cite{Allen:1993uc,Donadio:2010kp}. 
Both propagating  and diffusive modes provide significant contribution to $\kappa$, which depend on the diameter of the wire and the thickness the amorphous shell: for example for a wire with a diameter of 3 nm and an amorphous shell of 1 nm, propagating modes with frequencies up to 2 THz contribute about one half of the total $\kappa$ at room temperature.  
\begin{figure}[htb]
   \sidecaption
   \includegraphics[scale=.25]{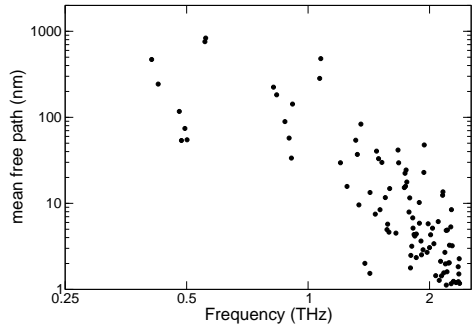} 
   \caption{Mean free paths of propagating phonons in a 2 nm thick silicon nanowire with few atoms thick amorphous shell.}
   \label{fig:NWmfp}       
\end{figure}

Given the delocalized character of both propagating and diffusive modes, which typically extend for several tens of nanometers, over the whole models considered in the simulation studies, one would argue that surface amorphization in thin wires deeply modifies the character of the vibrational modes, rather than just scattering phonons. 
It is interesting to point out that low frequency propagating phonons in rough wires can still exhibit remarkably long mean free paths of the order of 1 $\mu$m even in wires as thin as 2 nm (Figure~\ref{fig:NWmfp}), against the common wisdom that the diameter of the wire marks the upper limit of the mean free path of the heat carriers (Casimir limit). Very long phonon mean free paths were indeed recently observed in SiGe nanowires \cite{Hsiao:2013gz}.

Heat transport in thicker wires with diameters larger than 10 nm exhibit similar characteristics as those discussed for thin wires, however elastic scattering calculations on open systems and molecular dynamics simulations of periodic systems show that surface features become less effective in limiting $\kappa$ \cite{He:2012dn,Duchemin:2012ii}, and the microscopic origin of the reduction of $\kappa$ reported in some experiments remains debated. 
On the other hand, in thick wires mesoscopic models including surface scattering entail satisfactory results, provided that the correct phonon dispersion relations are used \cite{Mingo:2003ub}.  
Besides surface amorphization, several other possible strategies to reduce $\kappa$ in semiconducting wires were devised using atomistic simulations, such as introducing screw dislocations \cite{Xiong:2014de}, surface faceting~\cite{Sansoz:2011kc}, alloying \cite{Chan:2010ba}, and crystalline core-shell structuring \cite{Hu:2011bl}.
 
To conclude, it is worth mentioning that atomistic modeling predicted that suspended semiconducting wires would exhibit quantized thermal conductance  at low temperature in ballistic regime \cite{Rego:1998uh}. The quantum of conductance is universal, yet temperature-dependent, and corresponds to $\pi^2 k_B^2T/3h$. The possibility to probe the quantum of conductance is made viable by the harmonic phonon features described at the beginning of this session, which are general for suspended one-dimensional nanostructures. This prediction was verified experimentally few years later and quantized thermal conductance was measured in silicon nitride phonon waveguides \cite{Schwab:2000ub}.

\subsection{Ultra-thin Silicon Membranes}

Whereas suspended silicon nanowires display a high potential for tuning and optimization of the thermal conductivity that would be very attractive for thermoelectric energy conversion, their use in devices is complicated by their fragility and by the lack of scalable fabrication processes. In turn, quasi two-dimensional ultra thin membranes appear as more versatile systems for  applications that rely heavily on phononic properties, including sensors, nanomechanical resonators and thermoelectrics.  
Recent experiments demonstrated intriguing phononic and thermal properties of silicon membranes, including flexural acoustic modes with quadratic dispersion relations \cite{Cuffe:2012im}, quasi-ballistic transport at the micro scale at room temperature \cite{Johnson:2013ic}, and yet a significant reduction of thermal conductivity compared to bulk\cite{ChavezAngel:2014be}.
A compelling microscopic interpretation of these results can be obtained by atomistic modeling, which shows to what extent they emerge from the interplay of dimensionality reduction and surface features. 

\begin{figure}[htb]
   \sidecaption
   \includegraphics[scale=.26]{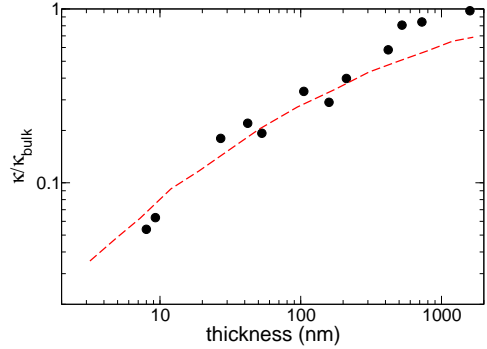} 
   \caption{Room temperature thermal conductivity of suspended silicon membranes normalized by the thermal conductivity of natural bulk silicon (148 W/m K at 300 K) \cite{ChavezAngel:2014be,Neogi:2015gk}. The dashed line is the result of a kinetic model with diffusive surface scattering \cite{Johnson:2013ic}.}
   \label{fig:membranes-exp}       
\end{figure}

The experimental measurements of $\kappa$ in thin films and suspended membranes can be interpreted by mesoscopic kinetic models~\cite{Callaway:1959uo} based on bulk phonon dispersion relations \cite{Johnson:2013ic}, in which surface scattering is modeled empirically, in analogy with the Sondheimer model for electronic transport \cite{Sondheimer:1952fz} (Figure~\ref{fig:membranes-exp}). However, comparing lattice dynamics results to molecular dynamics simulations it was shown that for films or membranes of the order of, or thinner than, $\sim 10$ nm large discrepancies may arise, thus indicating that models based on bulk properties cannot be used to predict the thermal conductivity of ultra-thin membranes \cite{Turney:2010ck}.

\begin{figure}[htb]
   \sidecaption
   \includegraphics[scale=.5]{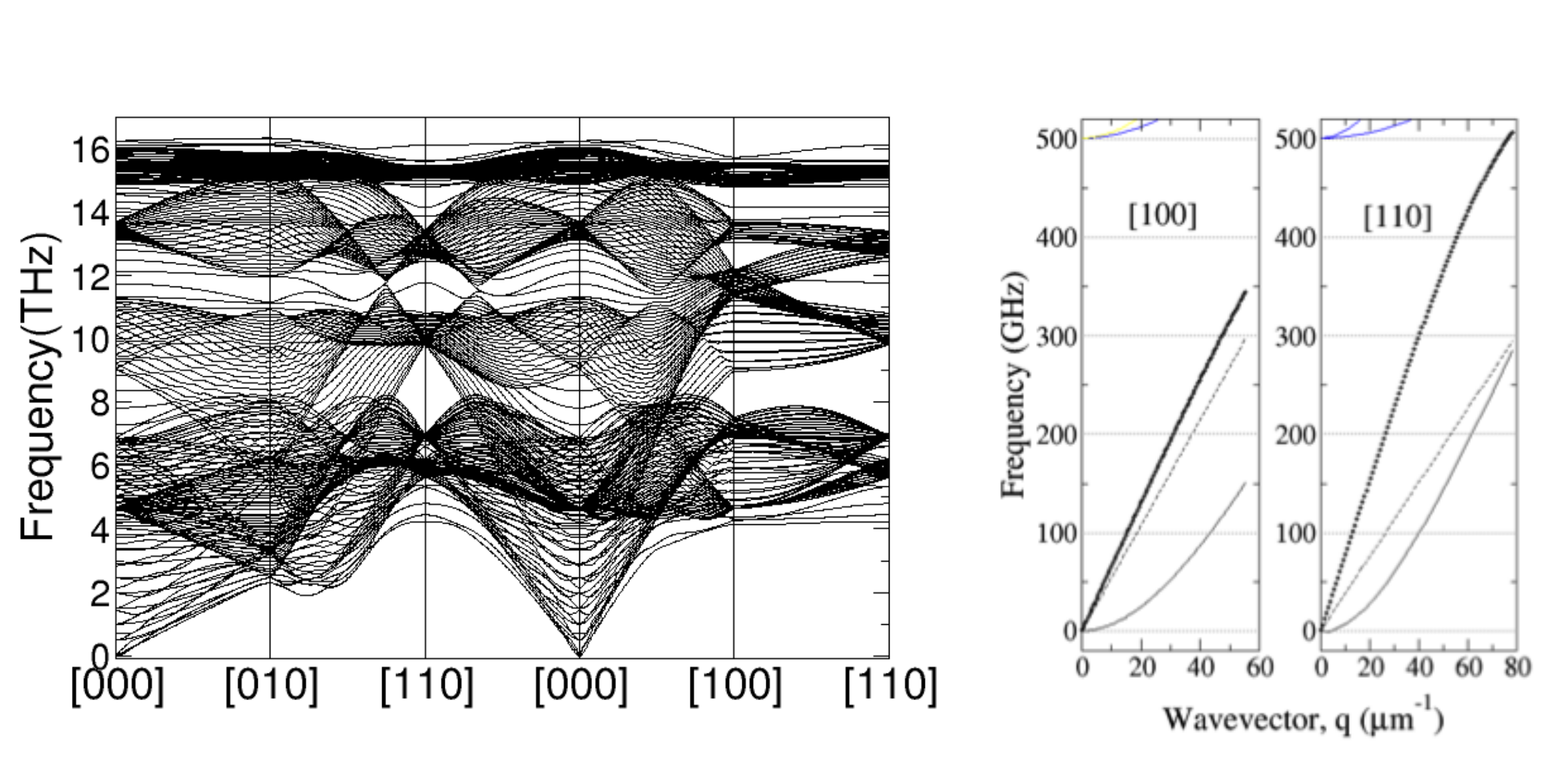} 
   \caption{Phonon dispersion relations of a 5 nm thick silicon membrane, and detail of the acoustic modes in proximity of the $\Gamma$ point \cite{Neogi:2015cp}.}
   \label{fig:memphonons}       
\end{figure}

\begin{figure}[htb]
   \sidecaption
   \includegraphics[scale=.38]{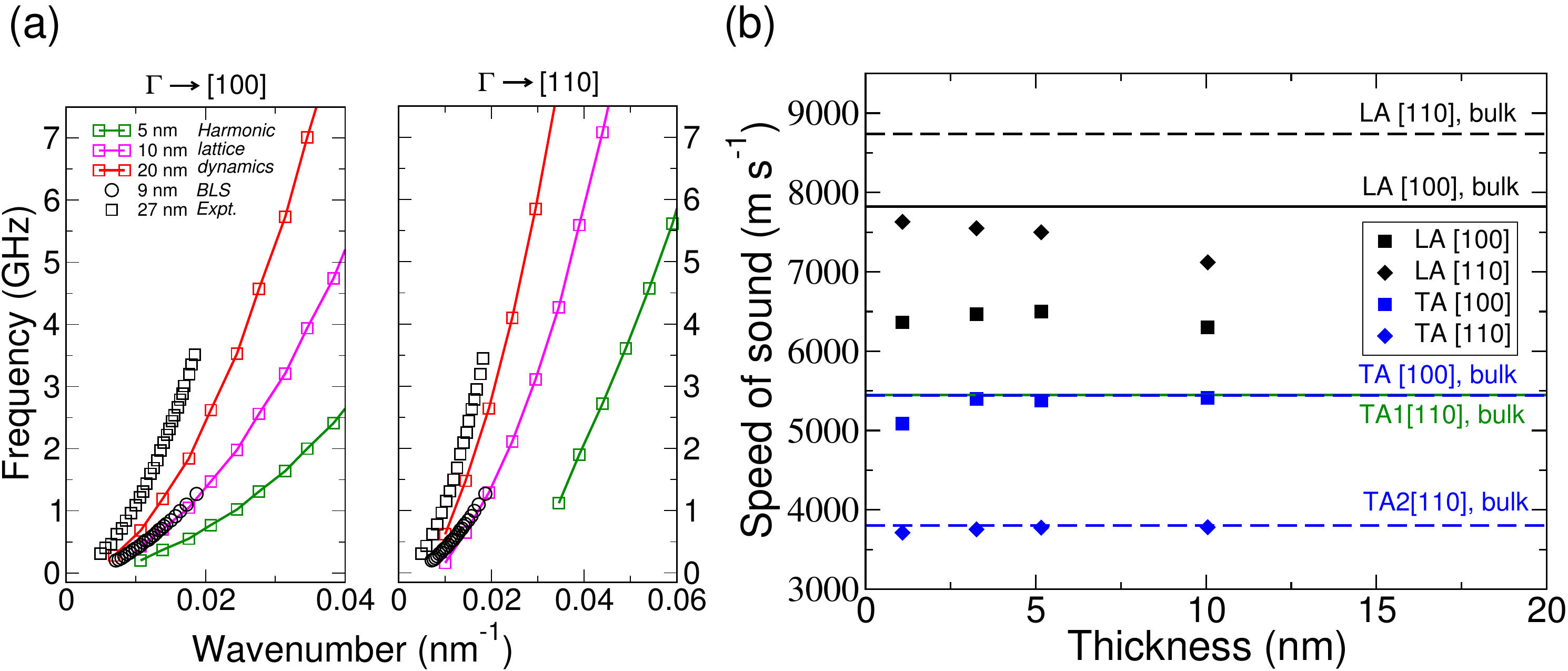} 
   \caption{Dispersion of the flexural modes of silicon membranes of different thickness (a), and dependence of the group velocities of in-plane longitudinal and transverse acoustic modes on membrane thickness (b). Data from Refs. \cite{Neogi:2015cp,Neogi:2015gk}.}
   \label{fig:flexmem}       
\end{figure}

As in the case of silicon nanowires discussed in the previous section, also for membranes it is necessary to assess the effect of dimensionality reduction on the harmonic and anharmonic phonon properties. 
If the interatomic interactions are modeled with inexpensive empirical potentials \cite{Tersoff:1989tr}, the phonon dispersion relations of atomistic models of silicon membranes with thickness up to few tens of nanometers can be calculated by diagonalization of  the dynamical matrix computed from Eq.~\ref{dynmat}. 
Lattice dynamics calculations of periodic crystalline models with ideal surface reconstruction allow us to asses the general effects of dimensionality reduction on phonons. Figure~\ref{fig:memphonons} reports the phonon dispersion relations of a 5 nm thick membrane.  
%
The large number of atoms in the unit cell gives rise to a large number of optical bands, which extend to low frequency and mix with the acoustic modes. The features of these bands, originated from removing periodicity in one direction, cannot be reproduced correctly by zone folding of bulk phonons, especially for the thinnest membranes, as they have derivative equal to zero at the $\Gamma$ point and $\omega\propto q^2$. 
The detail of the acoustic dispersions at the $\Gamma$ point shows that one of the transverse acoustic modes converts into a flexural mode with $\omega\propto q^2$.  
The flexural modes, related to out-of-plane vibration, are very sensitive to the thickness of the membrane: thinner membranes exhibit softer flexural modes (Figure~\ref{fig:flexmem}a). The dispersion relations of these modes was measured by Brillouin light scattering experiments, which showed a remarkable agreement with modeling \cite{Cuffe:2012im,Neogi:2015gk}
Also the speed of sound of the longitudinal acoustic branch depends on the thickness of the membranes, yet weakly. In general it is lower than in the bulk, but in the sub-10 nm regime the speed of sound of the LA modes decreases with thickness. On the other hand the speed of sound of the in-plane transverse modes, which remain linear in $q$, is not affected by variations of thickness (Figure~\ref{fig:flexmem}b).

\begin{figure}[htb]
   \sidecaption
   \includegraphics[scale=.34]{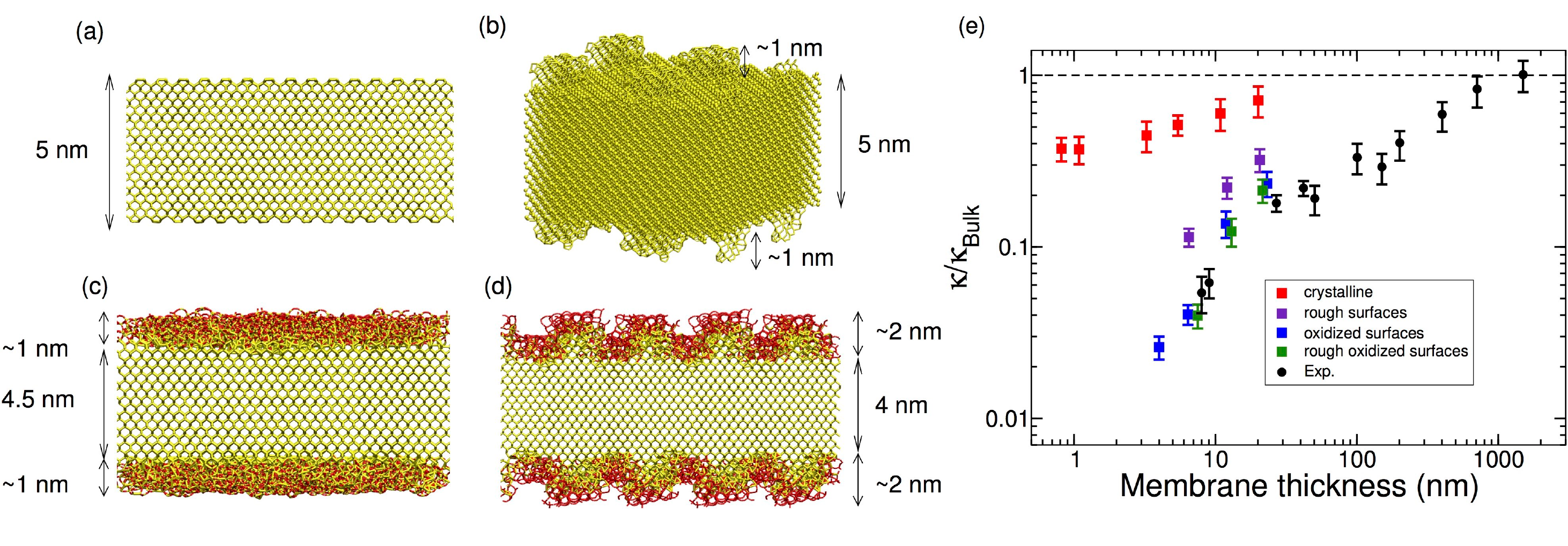} 
   \caption{Models of silicon membranes oriented in the (001) direction, with pristine crystalline surfaces (a), surface roughness (b), flat native oxide layers (c) and rough native oxide (d), and their thermal conductivity (e). Data from \cite{Neogi:2015cp,Neogi:2015gk}.}
   \label{fig:memkapp}       
\end{figure}

Whereas the characterization of the acoustic phonons from atomistic modeling exhibit a remarkable agreement with experiments,  the thermal conductivity of pristine crystalline membranes, computed by equilibrium MD, is in sharp disagreement with the measurements reported in \cite{ChavezAngel:2014be}. 
Crystalline membranes with ideal surfaces exhibit a reduction of $\kappa$ with respect to bulk silicon, however not as large as the one probed by experiments (Figure~\ref{fig:memkapp}). 
Nevertheless, as opposed to the case of nanowires, $\kappa$ decreases monotonically with decreasing thickness reaching $1/3$ of $\kappa_{bulk}$ for $\sim 1$ nm thick membranes. No evidence of possible divergence of $\kappa$ was observed.

The discrepancy with experiments indicates that pristine crystalline models are not representative for real systems.  
Since silicon membranes are exposed to air during and after fabrication, a layer of native oxide forms at their surfaces. Such layer is about 1 nm thick and can exhibit nanoscale roughness. 
When models with rough and/or oxidized surfaces as those shown in Figure~\ref{fig:memkapp}b-d are considered, the simulations reproduce the experimental $\kappa$ very well (Figure~\ref{fig:memkapp}e), thus indicating that also for ultra-thin silicon membranes surface properties dictate the major reduction of $\kappa$.
Recent experiments, in which the native oxide layer is removed by wet etching and eventually let re-grow, confirm the prominent role of surface oxidation and roughness on thermal transport. 
This effect is more appreciable in the thinnest samples measured ($\sim 8$ nm) for which an increase of $\kappa$ of about 2.3 times (from 8 to 18 Wm$^{-1}$K$^{-1}$) upon etching was observed. Further exposure of the same sample to air for several hours leads to a reduction of $\kappa$ to 12 Wm$^{-1}$K$^{-1}$. 
In thicker samples the same cycle of etching and re-oxidation lead to smaller variations of $\kappa$ \cite{Neogi:2015gk}. 

Non equilibrium MD simulations, which allow one to calculate the accumulation function of $\kappa$ as a function of phonon mean free path, show that surface scattering shifts the major contribution to the total thermal conductivity at room temperature from phonons with a broad range of mean free paths up to 1 $\mu$m for pristine crystalline membranes, to a much narrower distribution of mean free paths smaller 80 nm for membrane models with rough native oxide at the surfaces. 

Dimensionality reduction in membranes enables other approaches to control heat transport via surface modifications.
For example, it was suggested that drilling holes or depositing nanoscale pillars, of the same or of a different material. 
 would modify the band structure and enhance phonons scattering through local resonances, thus reducing the thermal conductivity \cite{Yu:2010fp,He:2011fh,Davis:2014hu,Graczykowski:2015du}. 

\section{Conclusions}
\label{sec:conclusions}

In this chapter we have illustrated a limited, yet representative, set of problems in which atomistic simulations are exploited to shed light on nanoscale heat transport in materials with reduced dimensionality. 
Even though models for specific materials are employed, common features emerge, stemming for example from symmetries and invariances associated with dimensionality reduction. 
Phonon dispersion relations and vibrational density of states are deeply affected by changes in dimensionality, especially concerning acoustic phonons. Specifically, the emergence of flexural phonons with quadratic dispersion relations sizably influences both sound and heat propagation in one and two-dimensional nanostructures. Flexural modes in 2D membranes and 1D wires are characteristic of low dimensional systems in a three-dimensional space, in which each atom has three degrees of freedom.
This aspect marks a major difference with non-linear low-dimensional systems, such as the Fermi-Pasta-Ulam model, in which motion is also confined in a space with reduced dimensionality. 
In addition, except for graphene, which is a truly atomically thick two-dimensional array of atoms, all the other systems known in nature have a complex three-dimensional structure, even though they may extend in one or two dimension only. 

Starting from this premises, it is then understandable that, at variance with non-linear models, no clear evidence of diverging thermal conductivity was found in atomistic simulations of nanostructures, even though both graphene and carbon nanotubes exhibit extremely high thermal conductivity, and diffusive (Fourier) heat transport has been predicted to occur only at macroscopic size scales. Phenomena like large ballistic phonon mean free path and the emergence of "second sound" at relatively high temperature, which lead to quasi-diverging $\kappa$ as in non-linear models, actually stem from the reduced phase-space available to anharmonic phonon scattering and the presence of flexural modes.

A further general consequence of dimensionality reduction is the very large surface-to-volume ratio, which make heat transport in low-dimensional nanostructures extremely sensitive to surface modifications, such as roughening, faceting, functionalization, oxidation and interactions with substrates. A possible analogy with statistical physics can be made considering the difference between momentum conserving and non-conserving models, such as the Frenkel-Kontorova model, where lack of momentum conservation, e.g from the interaction with an underlying fixed potential, suppresses divergence.
The tremendous impact of surfaces on nanoscale heat transport was predicted in many simulation studies, and probed in several	 experiments. These works have demonstrated that accurate modeling needs to include chemical specificity and benefits from direct feedback with experiments.

\bibliography{ultimate}
\end{document}